\documentclass[journal=jctcce,manuscript=article]{achemso}

\usepackage[version=3]{mhchem} % Formula subscripts using \ce{}

\author{Artur Lyssenko}
\affiliation[MIT]
{Department of Materials Science and Engineering, Massachusetts Institute of Technology, Cambridge, Massachusetts 02139, United States}
\alsoaffiliation[Second University]
{Department of Chemistry and Chemical Biology, Harvard University, Cambridge, MA, 02138, USA}

\author{KyuJung Jun}
\affiliation[MIT]
{Department of Materials Science and Engineering, Massachusetts Institute of Technology, Cambridge, Massachusetts 02139, United States}
\alsoaffiliation[Second University]
{School of Mechanical Engineering, Korea University, Seoul, South Korea, 02841}
\email{kyujung@korea.ac.kr}

\author{Rafael Gómez-Bombarelli}
\affiliation[MIT]
{Department of Materials Science and Engineering, Massachusetts Institute of Technology, Cambridge, Massachusetts 02139, United States}
\email{rafagb@mit.edu}
\affiliation[MIT]
{Department of Materials Science and Engineering, Massachusetts Institute of Technology, Cambridge, Massachusetts 02139, United States}

\title
  {Drude, Where's My Charge? End-to-End Learning of Polarizable Drude Force Fields for Polymer Electrolytes}

\begin{document}

%%%%%%%%%%%%%%%%%%%%%%%%%%%%%%%%%%%%%%%%%%%%%%%%%%%%%%%%%%%%%%%%%%%%%
%% The "tocentry" environment can be used to create an entry for the
%% graphical table of contents. It is given here as some journals
%% require that it is printed as part of the abstract page. It will
%% be automatically moved as appropriate.
%%%%%%%%%%%%%%%%%%%%%%%%%%%%%%%%%%%%%%%%%%%%%%%%%%%%%%%%%%%%%%%%%%%%%
% \begin{tocentry}

% Some journals require a graphical entry for the Table of Contents.
% This should be laid out ``print ready'' so that the sizing of the
% text is correct.

% Inside the \texttt{tocentry} environment, the font used is Helvetica
% 8\,pt, as required by \emph{Journal of the American Chemical
% Society}.

% The surrounding frame is 9\,cm by 3.5\,cm, which is the maximum
% permitted for  \emph{Journal of the American Chemical Society}
% graphical table of content entries. The box will not resize if the
% content is too big: instead it will overflow the edge of the box.

% This box and the associated title will always be printed on a
% separate page at the end of the document.

% \end{tocentry}

%%%%%%%%%%%%%%%%%%%%%%%%%%%%%%%%%%%%%%%%%%%%%%%%%%%%%%%%%%%%%%%%%%%%%
%% The abstract environment will automatically gobble the contents
%% if an abstract is not used by the target journal.
%%%%%%%%%%%%%%%%%%%%%%%%%%%%%%%%%%%%%%%%%%%%%%%%%%%%%%%%%%%%%%%%%%%%%
\begin{abstract}

Molecular dynamics simulations of solid polymer electrolytes require force fields that are fast enough to converge ion transport over hundreds of nanoseconds, yet accurately describe the ion–polymer interactions that govern it. Fixed-charge force fields meet the first requirement but neglect electronic polarization, an effect often approximated by scaling ionic charges using empirically chosen factors. Here we present AutoDrude, a differentiable framework that parametrizes polarizable Drude force fields directly from DFT energies, forces, and dipoles, with no empirical tuning and at only 1.20× the cost of a fixed-charge simulation. For PEO–LiTFSI, the dipole-constrained (hard-constraint) AutoDrude model reproduces the experimental ionic conductivity as accurately as the best charge-scaled model (17\% versus 16\% mean absolute percentage error). We compare the polarizable forcefield to fixed charge ones models by fitting a fixed charge model. Unlike the charge-scaled models tested here, however, AutoDrude also reproduces the dilute-limit cation transference number, its increase with salt concentration, and the absence of Li–TFSI contact ion pairs at low concentration. A penalty-based (soft-constraint) variant yields a closer fit to the DFT reference yet underestimates the conductivity by up to an order of magnitude, demonstrating that closer agreement with DFT training data alone does not guarantee accurate ion transport. These results indicate that although charge scaling can reproduce ionic conductivity the explicit inclusion of polarization is necessary to capture the concentration dependence of cation transport in PEO–LiTFSI.

\end{abstract}

%%%%%%%%%%%%%%%%%%%%%%%%%%%%%%%%%%%%%%%%%%%%%%%%%%%%%%%%%%%%%%%%%%%%%
%% Start the main part of the manuscript here.
%%%%%%%%%%%%%%%%%%%%%%%%%%%%%%%%%%%%%%%%%%%%%%%%%%%%%%%%%%%%%%%%%%%%%
\section{Introduction}

Solid polymer electrolytes (SPEs) are promising materials for next-generation lithium batteries because they offer intrinsic safety, enhanced mechanical robustness, and compatibility with high-energy lithium metal anodes.\cite{Li2023} Their performance is governed by the ability of the polymer matrix to transport lithium ions while maintaining electrochemical and thermal stability over long operating lifetimes. Because ion transport in SPEs is coupled to polymer segmental motion, the molecular-scale coordination environments of the ions ultimately dictate the macroscopic transport properties, and accurate microscopic modeling is therefore central to the design of improved SPE materials.\cite{Molinari2018PEO_LiTFSI,FranceLanord,Park2021,Dong2020} Poly(ethylene oxide) (PEO), whose ether oxygens strongly solvate Li$^+$, remains the most widely studied SPE host.\cite{Song2023Reflection}

Molecular dynamics (MD) simulations are widely used to probe the structural and dynamic origins of ion transport in polymer electrolytes.\cite{Molinari2018PEO_LiTFSI,FranceLanord} Ab initio and machine-learning interatomic potentials can accurately capture structural properties, but their cost restricts the length and time scales required for segmental relaxation and diffusive ion motion, so classical fixed-charge force fields remain the workhorse for simulations reaching hundreds of nanoseconds.\cite{Brooks2018} Because they neglect electronic polarization, fixed-charge force fields overbind Li ions to ether oxygens, overstabilize long-lived Li--anion contacts, and underestimate dielectric screening. These deficiencies are usually mitigated by empirically scaling the ionic charges against selected experimental observables.\cite{Liang2024,Gudla2020} As a result, many existing non-polarizable models of PEO-based electrolytes only partially reproduce mass densities, radial distribution functions (RDFs), ion cluster populations, ionic conductivity, and cation transference numbers.\cite{Brooks2018,Liang2024}

Explicit inclusion of polarization is widely recognized as a key requirement for predictive modeling of polymer electrolytes.\cite{Jing2019,MacKerell2016} In an electrolyte, the field around each ion is screened by the electronic response of the neighboring solvent and counterions, and the strength of this screening depends on the local coordination and on salt concentration. Fixed point charges cannot respond to their environment, so they overestimate ion--solvent and ion--ion attraction in the condensed phase. Uniform charge scaling, justified as a mean-field electronic continuum correction,\cite{Leontyev2011} restores the average screening but cannot follow local changes in coordination.\cite{Bedrov2019} Explicit polarization lets the effective interactions follow the local environment, which matters most for the properties that set transport in SPEs: ion pairing, cluster populations, and correlated ion motion.

Among polarizable models, the Drude oscillator formalism offers a computationally efficient representation of electronic response by attaching auxiliary charged particles to polarizable atoms via harmonic constraints.\cite{Lamoureux2003,MacKerell2016,Thole1981} In contrast to self-consistent induced-dipole approaches, which require iterative force evaluations at every timestep,\cite{Jing2019,Nocito2019,Shi2013,Borodin2006,Borodin2006b} Drude dynamics can be propagated using an extended Lagrangian formalism, retaining accuracy at a computational cost within roughly an order of magnitude of fixed charge force fields.\cite{Lamoureux2003,Yu2020,Lemkul2015} The combination of a physically motivated polarization model with GPU-accelerated MD engines (e.g., OpenMM) makes Drude-based simulations particularly attractive for SPE systems.\cite{Huang2021,Jiang2011,Lemkul2015}

Despite these advantages, parameterization of polarizable force fields for polymer electrolytes remains limited. Prior efforts have tuned specific subsets of parameters for select ion–polymer interactions\cite{Borodin2003,Borodin2006,Fang2021} or adapted small-molecule polarizable models with systematic optimization against high-level electronic structure data for selected atomic types.\cite{Kooi2020,Lemkul2021,MacKerell2016,Bedrov2019} Most notably, Borodin and Smith developed many-body polarizable force fields for ether- and carbonate-based electrolytes, including LiTFSI-doped polyethers,\cite{Borodin2006,Borodin2006b} and showed that MD with such a force field reproduces experimental ion self-diffusion coefficients and conductivity of PEO--LiTFSI melts.\cite{Borodin2006Macro} These force fields were parametrized fragment by fragment against quantum-chemistry data. AutoDrude instead fits all bonded, electrostatic, polarization, and screening parameters simultaneously and end-to-end.

Here, we introduce a machine-learning framework for fitting Drude polarizable force fields which we call AutoDrude. This framework builds on AutoBADDIE, a prior work developed to fit class 1 and class 2 interatomic potentials.\cite{Leon2025} We demonstrate AutoDrude on PEO with lithium bis(trifluoromethanesulfonyl)imide (LiTFSI), the prototypical SPE model system owing to its well-characterized coordination chemistry and experimental tractability.\cite{Song2023Reflection} Despite decades of progress, PEO--LiTFSI electrolytes still exhibit limited ionic conductivity and low cation transference numbers, especially at high salt concentrations where ion pairing and aggregation become prominent,\cite{Pesko_2017,Molinari2018PEO_LiTFSI} and it is precisely these structural and kinetic properties that fixed-charge models struggle to capture. Our training framework simultaneously optimizes bonded, electrostatic, polarization, and Thole screening parameters, using density functional theory (DFT) energies and forces, together with molecular dipole moments evaluated from the DFT-derived CHELPG point charges, as reference data. We develop a fully differentiable Drude-relaxation procedure that eliminates the need for costly self-consistent MD evaluations during training, enabling efficient optimization over large datasets. The resulting model is exported in CHARMM format for use in standard MD engines.

The resulting force field is used to evaluate agreement between simulation and experimental measurements with a focus on structural and kinetic properties. In addition, we fit a non-polarizable force field within the same framework and apply a range of charge scaling factors to systematically benchmark the impact of reduced-charge models on structural and kinetic properties.

\section{Methods}

\begin{figure}

    \includegraphics[width=\linewidth]{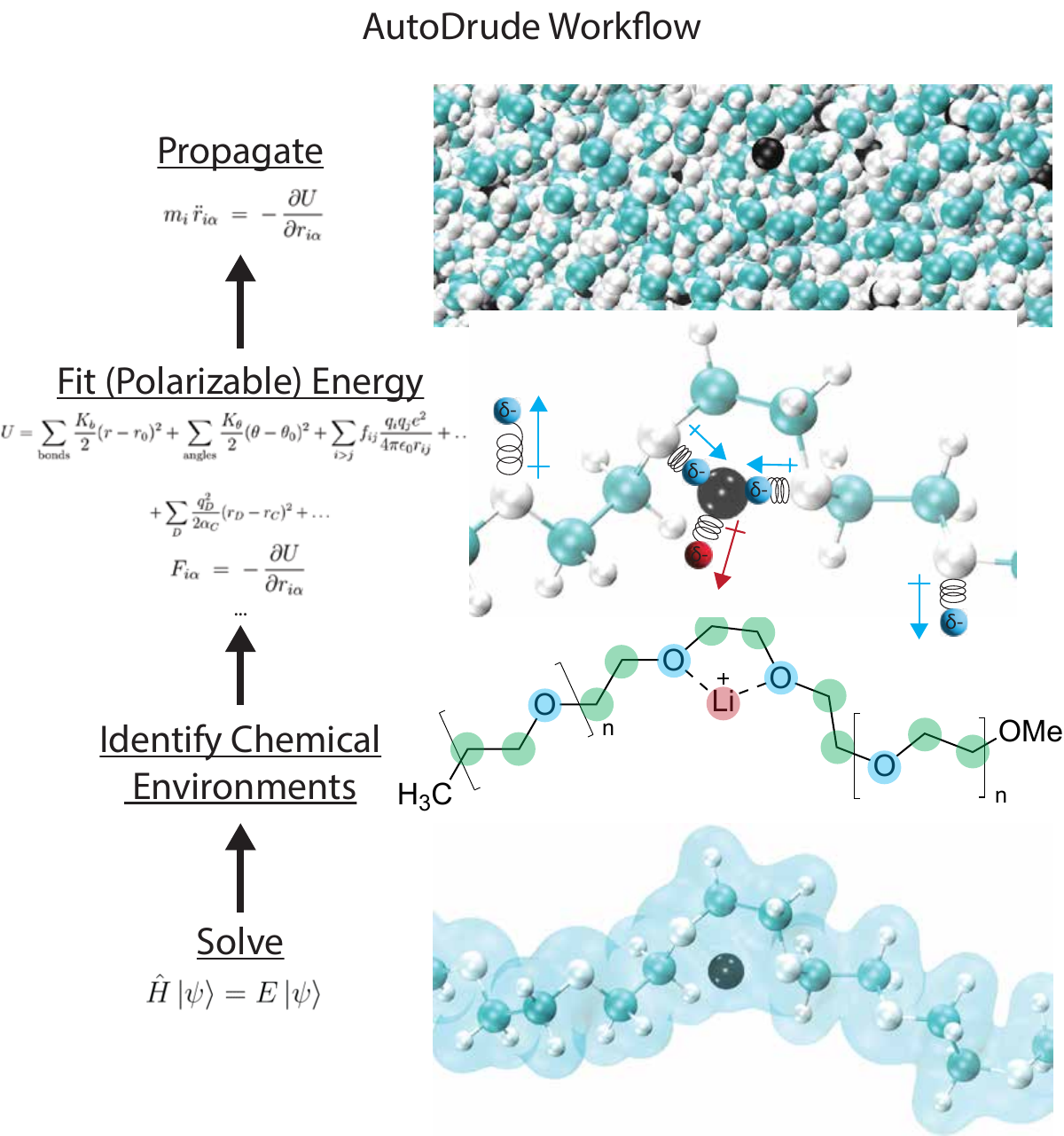}

    \caption{Schematic of the bottom-up steps to fit the force field as well as compute structural and transport properties.}
    \label{fig:schematic}
\end{figure}

A high-level schematic of the workflow is shown in Fig.~\ref{fig:schematic}.

\subsection{Polarizable Energy Function}
We adopt the CHARMM functional form for the underlying force field. A detailed description and justification of this functional form can be found elsewhere.\cite{Vanommeslaeghe2010}

\begin{equation*}
\begin{split}
        U_{CHARMM} &= \sum_{bonds}\frac{K_b}{2}(r-r_0)^2+\sum_{angles}\frac{K_{\theta}}{2}(\theta-\theta_0)^2+\sum_{dihedrals} \sum_{i=1}^4\frac{V_{i}}{2}(1 + \cos(i\phi- \phi_{0_i})) \\
        &+\sum_{i>j}f_{ij}(\frac{A_{ij}}{r^{12}_{ij}}-\frac{C_{ij}}{r^6_{ij}}+ \frac{q_i q_j e^2}{4\pi \epsilon_0 r_{ij}})
\end{split}
\end{equation*}

Electronic polarization is incorporated using the classical Drude oscillator formalism.\cite{Huang2021, Thole1981} In this approach, each polarizable atom (center) is attached to a mass-bearing auxiliary particle (Drude particle) via a harmonic spring. The Drude particle carries a charge $q_D$ and is harmonically restrained to its host center with a force constant $k_D$. The force constant ($k_D$) is related to the atomic polarizability $\alpha_C$ through $k_D = \frac{q_D^2}{\alpha_C}$.

Drude particles interact electrostatically with other Drude particles and atoms. Short-range dipole-dipole interactions between bonded (1-2 and 1-3) Drude pairs are screened using a Thole screening function $S_{ij}(r_{ij})$ which prevents unphysical polarization catastrophes at small separations.\cite{Thole1981, Huang2021} As in the standard CHARMM Drude force field, Coulomb interactions between 1-2 and 1-3 bonded atom pairs are excluded, while the corresponding Drude-Drude and Drude-atom dipole interactions are retained and Thole-screened. Interactions involving non-polarizable atoms are not screened. The Thole screening function is given by:

$$S_{ij}(r_{ij})  = 1- (1+\frac{a_i+a_j}{2(\alpha_i\alpha_j)^{1/6}}r_{ij})e^{-\frac{a_i+a_j}{(\alpha_i\alpha_j)^{1/6}}r_{ij}}$$

The total potential energy is therefore the sum of the CHARMM bonded and non-bonded terms, Coulomb interactions involving Drude particles, and the harmonic Drude spring energy.

The Drude degrees of freedom are propagated with an extended Lagrangian approach.\cite{Lemkul2015} In this formulation, a small portion of mass from the host atom is assigned to the Drude particle (0.4 amu in this work), allowing the Drude coordinates to be integrated dynamically alongside nuclear degrees of freedom. The equations of motion are integrated in a transformed center of mass frame, and the self-consistent-field (SCF) potential energy surface (PES) is well approximated on average as long as the Drude particles are thermostatted to a low temperature (1 K).\cite{Lemkul2015} This approach avoids explicit SCF iterations while retaining an accurate representation of electronic response at a computational cost significantly lower than induced-dipole methods.

\subsection{Data and Parameter Initialization}
To train our models, reference energies, forces, charges, and dipoles were computed using ORCA\cite{Neese2020} with the $\omega$B97X-D3 functional\cite{Lin2013} and the def2-TZVP basis set.\cite{Weigend2005} Atomic charges were obtained according to the CHELPG partitioning scheme.\cite{Breneman1990} The molecular geometries used for training were generated in a previous study.\cite{Leon2025} Lennard--Jones parameters were fixed to established CHARMM polarizable force-field values (Supporting Information S1). The dataset comprises 5,942 DFT-labeled configurations of seven species: isolated triglyme (3,172 frames from xTB bond, angle, and dihedral scans, classical MD, and xTB Hessian displacements) and tetraglyme (520 frames from classical and MLIP MD); Li--TFSI pairs (1,539 frames from classical MD, DFT Hessian displacements, and pose generation); and TFSI--TFSI (274), tetraglyme--Li (153), tetraglyme--TFSI (145), and tetraglyme--TFSI--Li (139) clusters sampled with MLIP MD (Supporting Information Figure~\ref{fig:Dataset}). Within each species, 10\% of the frames were held out at random as a test set and the remaining 90\% were used for training with no separate validation set was used. A detailed description of the geometry generation is provided in the Supporting Information S1.

\subsection{Drude Equilibration}
As Drude particles are auxiliary degrees of freedom, their equilibrium positions are not defined in our quantum mechanical training data. A straightforward strategy would be to perform a Drude relaxation using an external molecular dynamics engine at each training iteration. However, such engines are not differentiable, preventing backpropagation through the relaxation procedure. As a result, the regressor would receive limited information on how Drude parameters affect relaxed positions and the procedure would become computationally prohibitive for large datasets. 

To address this, we implement a fully differentiable Drude relaxation in PyTorch. We explore two optimization strategies: a hard constraint procedure which enforces the molecular dipoles as an equality constraint via a constrained optimization and a soft constraint procedure which includes an additional dipole penalty term in the loss function. In both cases, gradient descent is first used to find the local minimum for the Drude positional degrees of freedom. The dipole reported in the further discussion is computed from combined charges of the Drude particles and centers on the nuclear coordinates.

For the hard constraint we define a Lagrangian for each geometry $f$ that enforces agreement between predicted and reference molecular dipoles.

\begin{equation*}
    \begin{split}
        L_f = U_f + \sum_\alpha \lambda_{f\alpha} (p_{f\alpha} -\hat{p}_{f\alpha})
    \end{split}
\end{equation*}

where $U_f$ is the potential energy, $p_{f\alpha}$ is the predicted dipole moment, $\hat{p}_{f\alpha}$ is the reference dipole, and $ \lambda_{f\alpha}$ are Lagrange multipliers. Latin indices denote Drude particles and Greek indices denote Cartesian components. We then seek the stationary point of $L_f$ with respect to both the Drude positions ($x_{fj\beta}$) and the multipliers ($\lambda_{f\alpha}$),

\begin{equation*}
    \begin{split}
        \frac{\partial L_f}{\partial x_{fj\beta}} &= -F_{fj\beta} + \lambda_{f\beta}\,q_j = 0, \\
        \frac{\partial L_f}{\partial \lambda_{f\alpha}} &= p_{f\alpha} - \hat{p}_{f\alpha} = 0,
    \end{split}
\end{equation*}

where $F_{fj\beta}$ are the forces on the Drude particles and $q_j$ are their charges. The first line uses $\partial p_{f\alpha}/\partial x_{fj\beta} = q_j\,\delta_{\alpha\beta}$, which follows because the molecular dipole is linear in the Drude coordinates.

We solve the resulting Karush-Kuhn-Tucker (KKT) system\cite{Nocedal2006} with Newton's method:

\begin{equation*}
    \begin{bmatrix}
        H_{f,i\alpha\, j\beta} & q_i\,\delta_{\alpha\gamma}\\
        q_j\,\delta_{\gamma\beta} & 0
    \end{bmatrix}
    \begin{bmatrix}
    \Delta x_{fj\beta} \\
    \Delta \lambda_{f\gamma}
    \end{bmatrix}
    =
        -\begin{bmatrix}
        -F_{fi\alpha} + \lambda_{f\alpha}\,q_i\\
         p_{f\gamma} - \hat{p}_{f\gamma}
    \end{bmatrix}
\end{equation*}

where $H_{f,i\alpha\,j\beta} = \partial^2 U_f/\partial x_{fi\alpha}\,\partial x_{fj\beta} = -\partial F_{fi\alpha}/\partial x_{fj\beta}$ is the Hessian of the potential energy with respect to the Drude coordinates, obtained by forward-mode automatic differentiation of the Drude force function, and $\Delta\lambda_{f\gamma}$ are the updates to the three Lagrange multipliers of geometry $f$. A small diagonal regularization ($10^{-6}$) is added to $H$ for numerical stability.

Rows of the first block are indexed by $(i,\alpha)$ and rows of the second by $\gamma$. Summation over the repeated column indices $(j,\beta)$ and $\gamma$ is implied. The unknown vector for each geometry has $3N_D + 3$ components.

In practice, a few gradient-descent warm-up steps are followed by Newton iterations. The number of steps required is reported in the Results.

\subsection{Parameter Reduction}

Due to the repeated structure of polymers it is possible to type the resulting atoms in a way where only the monomer has free parameters (see 'Identify Chemical Environments' Fig. \ref{fig:schematic}). This drastically reduces the number of distinct parameters and improves data efficiency.

We adopt the graph-based message passing algorithm introduced in ref.~\citenum{Leon2025}, which differentiates atomic environments based on their bonded connectivity. We use a neighborhood depth of three bonds to respect the Thole screening on bonded 1-2/1-3 dipoles.

\subsection{Loss Function}

The total loss consists of weighted energy and force terms with physically motivated regularization. Consistent with previous work, we use L1 regularization on dihedral parameters, atomic charge constraints to match CHELPG charges, and molecular charge constraints to match their formal charge.\cite{Leon2025} We find it necessary to constrain the equilibrium bond lengths and angles near their average DFT values from the dataset to prevent nonphysical parameter drift during training. This mirrors prior polarizable force field development strategies in which equilibrium geometries (bond length and angles) were fixed to DFT-optimized geometries.\cite{Borodin2003} We also regularize the Drude spring constant toward $1000 \frac{\text{kcal}}{\text{mol·\AA}^2}$ consistent with the literature.\cite{MacKerell2016} The relative weights of the energy, force, and dipole terms ($\lambda_{\text{E}}$, $\lambda_F$, and $\lambda_{\mu}$) were set to $\lambda_{\text{E}} = 10$, $\lambda_F = 1$, and $\lambda_{\mu} = 10$ for the soft-constraint model ($\lambda_{\mu}$ effectively zero for the hard-constraint model, whose dipoles are enforced exactly, and for the non-polarizable model), based on a small number of trial trainings ($\lambda_{\text{E}} \in \{1, 10\}$; $\lambda_{\mu} \in \{10^{-8}, 1, 10, 10^{2}\}$) rather than a systematic search, and the parity plots (Fig.~\ref{fig:main-parity}) show the training and held-out test sets.. The remaining regularization weights were tuned so that the fitted parameters remained physical: all bond force constants positive, all angle force constants non-negative, all Thole factors positive, Drude spring constants within 1.1\% of 1000~kcal\,mol$^{-1}$\,\AA$^{-2}$, and equilibrium bond lengths and angles within 0.09~\AA\ and 3.7$^\circ$ of their DFT averages. The dipole term appears only for the soft constraint case.

$$\begin{aligned}
L &= \lambda_{\text{E}}\frac{1}{N_{\text{atoms}}} 
    \left\lVert \vec{E}_f - \vec{E}_{f,\text{True}} \right\rVert_2
    + \lambda_F \frac{1}{3N_{\text{atoms}}} 
    \left\lVert \vec{F}_{f i \alpha} - \vec{F}_{\text{True}} \right\rVert_2 \\[6pt]
  &\quad + \lambda_{\text{dih}} \sum_{\text{dih}} \left| \text{dih}_{\text{par}} \right|
    + \lambda_c \frac{1}{N_{\text{atoms}}} 
      \sum_i \left( q_{f i} - q_{\text{DFT}} \right)^2 \\[6pt]
  &\quad + \lambda_{\text{mol}} 
      \sum_i \left( q_{\text{mol}} - q_{\text{mol,FC}} \right)^2
    + \lambda_{\text{bond}} 
      \sum_i \left( r_0 - r_{0,\text{DFT}} \right)^2 \\[6pt]
  &\quad + \lambda_{\text{angle}} 
      \sum_i \left( \theta_0 - \theta_{0,\text{DFT}} \right)^2
    + \lambda_D \frac{1}{N_D} 
      \sum_i \left( 1000 \frac{\text{kcal}}{\text{mol·\AA}^2} - k_D \right)^2 \\[6pt]
    &\quad + \lambda_{\mu}\frac{1}{N_f}\sum_f \lVert \vec{\mu}_f - \vec{\mu}_{f,\text{ref}}\rVert^2
\end{aligned}$$

\subsection{Molecular Dynamics}

All molecular dynamics simulations were performed using the OpenMM package.\cite{Eastman2024} Each system consisted of ten polymer chains of 128 ethylene oxide repeat units with 13, 39, 96, or 153 LiTFSI ion pairs (0.23, 0.68, 1.7, and 2.7~molal; $r = [\mathrm{Li}]/[\mathrm{EO}] = 0.010$, 0.030, 0.075, and 0.120). All simulations were conducted at 353~K. Long-range electrostatics were treated using particle-mesh Ewald (PME) with a real-space cutoff of 1.2 nm and an Ewald error tolerance of 0.0001. Lennard-Jones interactions employed a switching function beginning at 1.0 nm with a cutoff of 1.2 nm. Bond lengths involving hydrogen atoms were constrained using the SHAKE algorithm. Pressure was maintained using a Monte Carlo barostat.

Polarizable simulations employed the DrudeLangevinIntegrator with friction coefficients of 5 ps$^{-1}$ and 20 ps$^{-1}$ for the atomic and Drude degrees of freedom. Drude particles were thermostatted at 1 K, and a hard-wall constraint of 0.02 nm was imposed on the Drude-center separation to prevent polarization catastrophe. Non-polarizable simulations were performed using the LangevinIntegrator with a friction coefficient of 5 ps$^{-1}$.\cite{Huang2021}

Initial configurations were generated using PACKMOL. \cite{Martinez2009} The equilibration protocol follows ref.~\citenum{Ruza2025} with minor modifications for polarizable simulations. For Drude systems, equilibration proceeded as follows: (i) 50 ps in the NVT ensemble at 50 K with a timestep of 0.5 fs to allow the Drude degrees of freedom to relax, (ii) 100 ps in the NPT ensemble with pressure ramped from 1 to 100 bar using a timestep of 0.5 fs, (iii) 100 ps of heating in the NPT ensemble to the target temperature (353 K) using a timestep of 0.25 fs. Non-polarizable simulations followed the protocol described in ref.~\citenum{Ruza2025}, using a 1 fs timestep. All production simulations were carried out for 300 ns with a timestep of 1~fs.

\subsection{Transport Analysis}
Ionic conductivities and cation transference numbers were computed using the cluster Nernst--Einstein (cNE) approach.\cite{FranceLanord} Wheeler--Newman (Onsager) conductivities evaluated directly from the collective ion displacements were not converged within 300~ns for the present system size, and we therefore report cNE values, which converge faster because they depend only on single-cluster diffusion. The reported uncertainties reflect replica-to-replica scatter only. System-size dependence was not examined.

\section{Results and Discussion}

\subsection{Differentiable Drude Relaxation}
The differentiable relaxation converges in a handful of steps, which makes it practical inside the training loop. For the soft constraint, ten gradient-descent steps suffice for the Drude positions, dipoles, and forces to converge. For the hard constraint, three gradient-descent warm-up steps followed by three Newton steps on the KKT system suffice, and none of the convergence metrics change appreciably between three and five Newton steps (Supporting Information S2). Because every step is differentiable, gradients of the loss propagate through the relaxation to all Drude parameters, including the Thole screening factors that enter only through an exponential.

\subsection{Training}

The final models place Drude particles on oxygen, nitrogen, and lithium atoms only, and this configuration was used for all MD simulations. We arrived at it by removing polarization top-down, starting from all non-hydrogen atoms: models that also polarized carbon or sulfur trained unstably (degraded energy and force parity for the hard constraint and unphysical negative Thole parameters for the soft constraint), and models that additionally polarized fluorine trained stably, but the intramolecular Thole factor of fluorine collapsed to zero ($\leq 10^{-3}$, from an initial 1.3) in both variants, and an MD test run at 1.7~molal showed artificial salt segregation: Li$^+$ remained poorly solvated by PEO (1.6 ether oxygens within 2.8~\AA, versus 4.9 for the final model) and the ion density became spatially non-uniform. Parity plots and a stability analysis for each intermediate configuration are provided in the Supporting Information S3. All models were trained for up to 2000 epochs.

\begin{figure}

    \includegraphics[width=\linewidth]{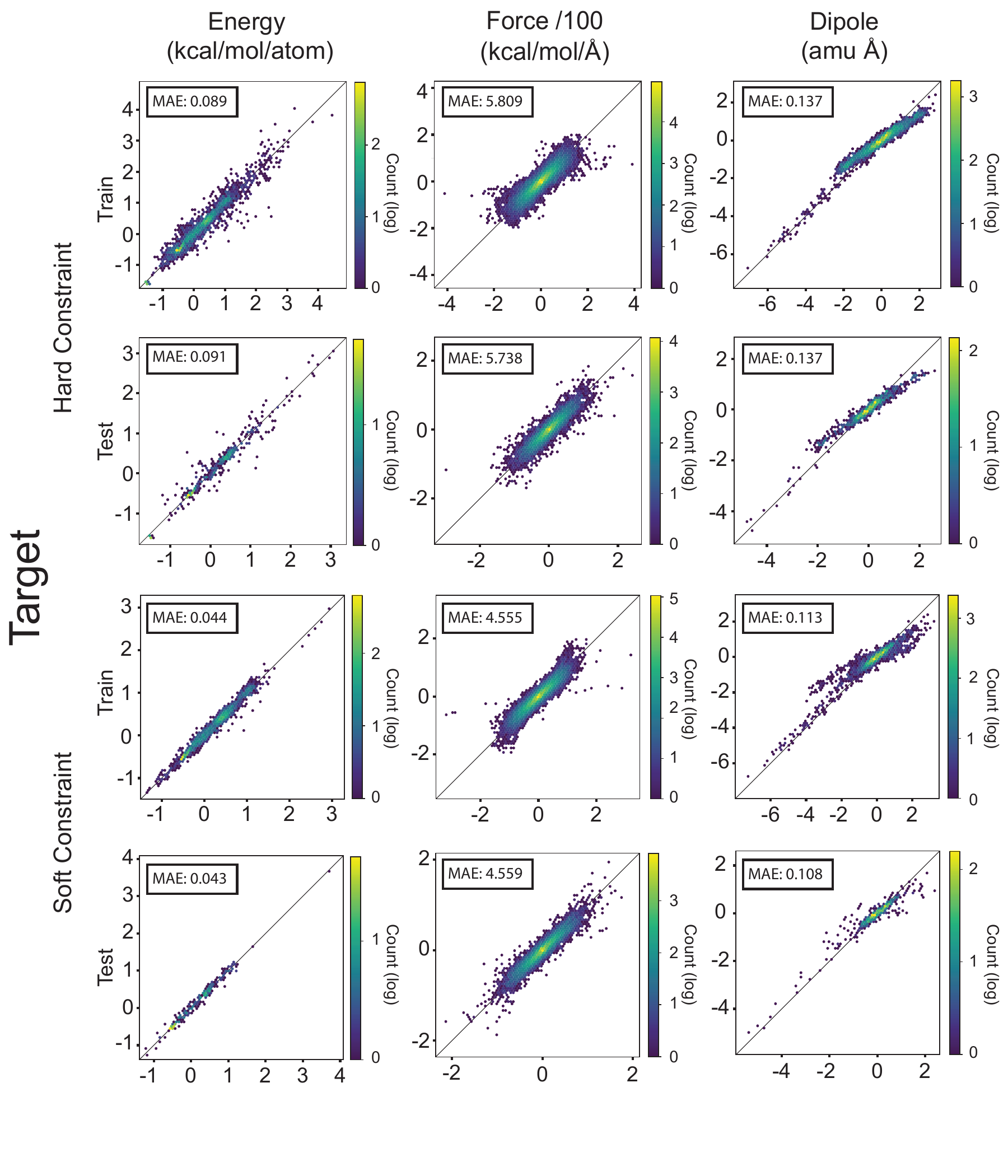}

    \caption{Parity plots for the training and test sets for both hard- and soft-constraint algorithms. Predicted values are shown on the x-axis and DFT reference values on the y axis.}
    \label{fig:main-parity}
\end{figure}

As seen in the first two columns of Figure~\ref{fig:main-parity}, the soft constraint algorithm achieves lower mean absolute errors (MAE) for both energy and force predictions relative to DFT. We attribute this to the greater flexibility in exploration of the parameter space as Drude positions are not strictly constrained to lie along the molecular dipole axis. This is further supported by the dipole parity plots in the last column of Figure~\ref{fig:main-parity}. The soft constraint algorithm yields a lower dipole MAE but the hard constraint algorithm exhibits reduced scatter and a stronger clustering of points along the parity line. This indicates improved reproduction of dipole orientations and magnitudes across geometries, despite slightly higher average error compared to the soft-constraint formulation.

\subsection{Property Validation}

We first compare the relative energies, $E - E_\text{min}$, of 103 Li--TFSI configurations optimized in vacuum at the $\omega$B97X-D3/def2-TZVP level, none of which is part of the training set. For each configuration, the local coordination environment of Li-ion was classified based on its two nearest coordinating atoms. We distinguish between (i) oxygen-nitrogen coordination, (ii) two oxygen atoms bound to the same sulfur center, and (iii) two oxygen atoms bound to different sulfur centers within TFSI.

\begin{figure}[tbp]

  \includegraphics[width=\linewidth]{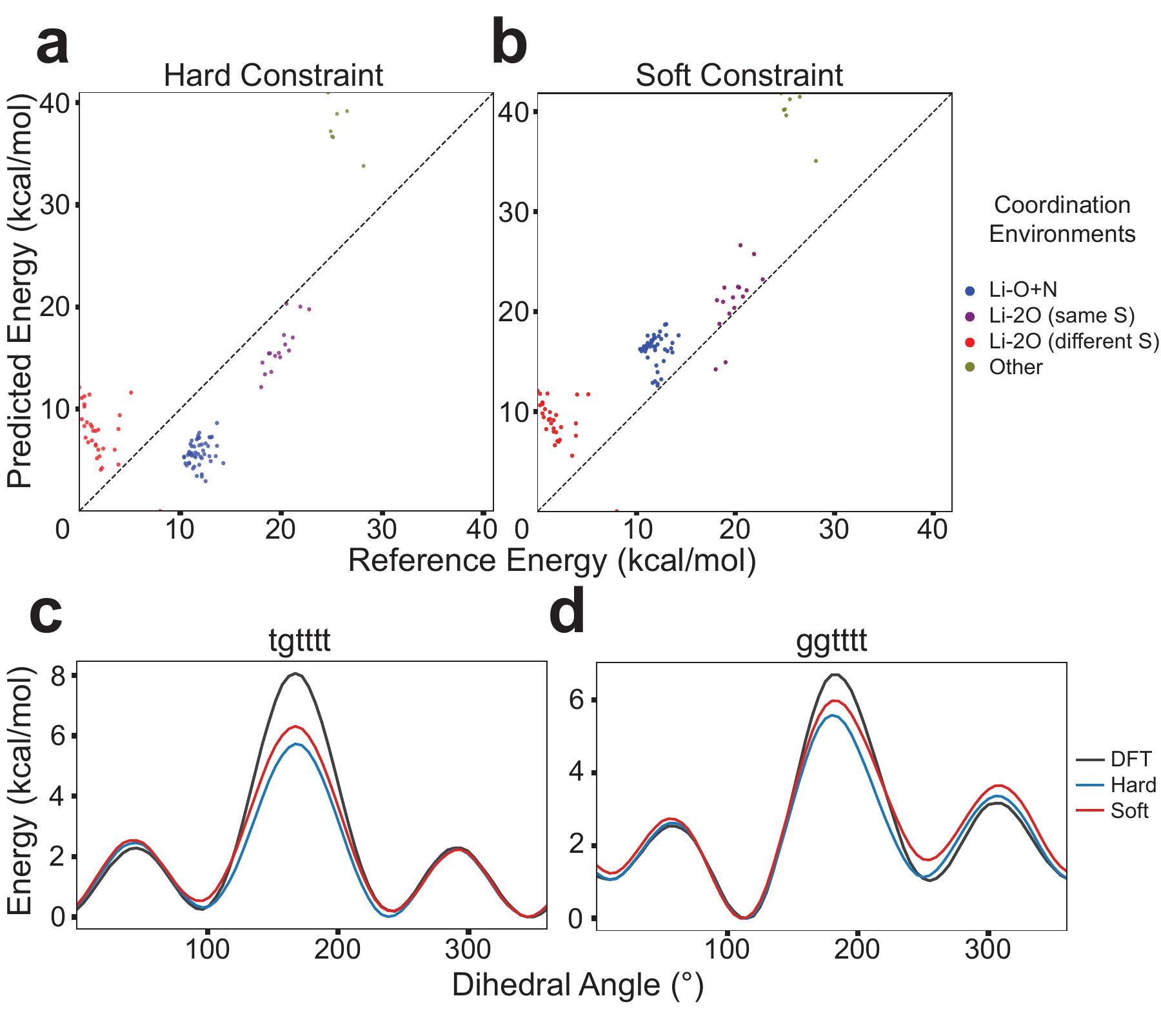}
  \caption{(a) and (b) Predicted versus DFT relative energies, $E - E_\text{min}$, of 103 Li--TFSI configurations not included in training. Coordination environments are based on the two nearest neighbors of the lithium-ion. Oxygen atoms bound to different sulfur centers within TFSI are distinguished from those bound to the same sulfur. (c-d) Dihedral scans of diglyme comparing DFT and fitted force-field energy profiles. The characters “t” and “g” denote trans or gauche conformations, with the right-most character indicating the scanned dihedral.}
  \label{fig:tfsi-coord-dih}
\end{figure}

As shown in Fig.~\ref{fig:tfsi-coord-dih}(a-b), the soft constraint algorithm more accurately captures the relative energetic ordering of these coordination motifs compared to DFT. The hard constraint model systematically overstabilizes the oxygen-nitrogen coordination (i) and two oxygen coordination from the same sulfur center (ii) whereas the soft constraint model overestimates the energies of all coordination environments. These results suggest that greater flexibility in the soft-constraint algorithm improves the relative description of local ion coordination energetics. Evaluated on the same 103 configurations, the non-polarizable model reproduces the energetic ordering of the motifs at least as well (Spearman $\rho = 0.89$, versus 0.84 and 0.35 for the soft- and hard-constraint models) but has the largest absolute error (MAE 6.4 versus 5.7 and 6.2~kcal\,mol$^{-1}$), overestimating the oxygen nitrogen motif by 8.5~kcal\,mol$^{-1}$ on average (Supporting Information Figure~\ref{fig:classical-poses-scans}).

We further assess the intramolecular torsional energetics by performing constrained dihedral scans of diglyme, a molecule not included in the training set, at the $\omega$B97X-D3/def2-TZVP level. Figure~\ref{fig:tfsi-coord-dih}(c-d) compare DFT energy profiles with those predicted by the fitted force fields. In the notation shown, 't' and 'g' denote trans and gauche conformations, respectively, with the rightmost letter indicating the dihedral scanned. Both force-field variants reproduce the positions of the torsional minima and barriers but underestimate the central barrier height, most visibly for tgtttt. Over all ten scans (722 points), the mean absolute errors are 0.56 and 0.51~kcal\,mol$^{-1}$ for the hard- and soft-constraint models, respectively. The non-polarizable model, evaluated on the same scans, performs comparably (0.48~kcal\,mol$^{-1}$) and also underestimates the tgtttt barrier compared to DFT (7.0 versus 8.1~kcal\,mol$^{-1}$ Supporting Information Figure~\ref{fig:classical-poses-scans}). Additional dihedral scans are provided in the Supporting Information Figure~\ref{fig:alldih}. 

\subsection{Structural Analysis}

To validate the structural properties of the simulated PEO--LiTFSI electrolytes, we first computed the mass density at 353~K as a function of salt concentration (Table~\ref{tab:density_constraints}). Both polarizable models show the expected monotonic densification upon salt addition, with the soft-constraint model slightly denser at every concentration. Compared with densities measured at 363~K for 5~kg\,mol$^{-1}$ PEO/LiTFSI,\cite{Pesko_2017} both polarizable models reproduce the trend but underestimate the magnitude, by 6--9\% for the hard-constraint and 3--8\% for the soft-constraint model. The charge-scaled non-polarizable models deviate by a similar 3--9\% (Table~\ref{tab:density_scaling_no_pol}). Because the deficit is common to polarizable and non-polarizable models, it suggests an origin in the fixed Lennard--Jones parameters rather than in the treatment of polarization. The reported uncertainties reflect only the spread between two independent simulations while systematic errors from the force field, including the fixed Lennard--Jones parameters, and from comparing against measurements at other temperatures are likely larger.

\begin{table}[h!]
\centering
\caption{Simulated densities (g\,cm$^{-3}$) of PEO--LiTFSI systems at 353~K and various molalities for the hard and soft constraint models. Values are means of two independent simulations. Their standard deviation is below 0.01~g\,cm$^{-3}$ in all cases.}
\begin{tabular}{lcccc}
\hline
\textbf{Molality (mol\,kg$^{-1}$)} & 0.23 & 0.68 & 1.7 & 2.7 \\
\hline
\textbf{Hard constraint} & 1.06 & 1.12 & 1.21 & 1.28 \\[2pt]
\textbf{Soft constraint} & 1.07 & 1.13 & 1.26 & 1.34 \\
\hline
\end{tabular}
\label{tab:density_constraints}
\end{table}

\begin{figure}

  \includegraphics[width=\linewidth]{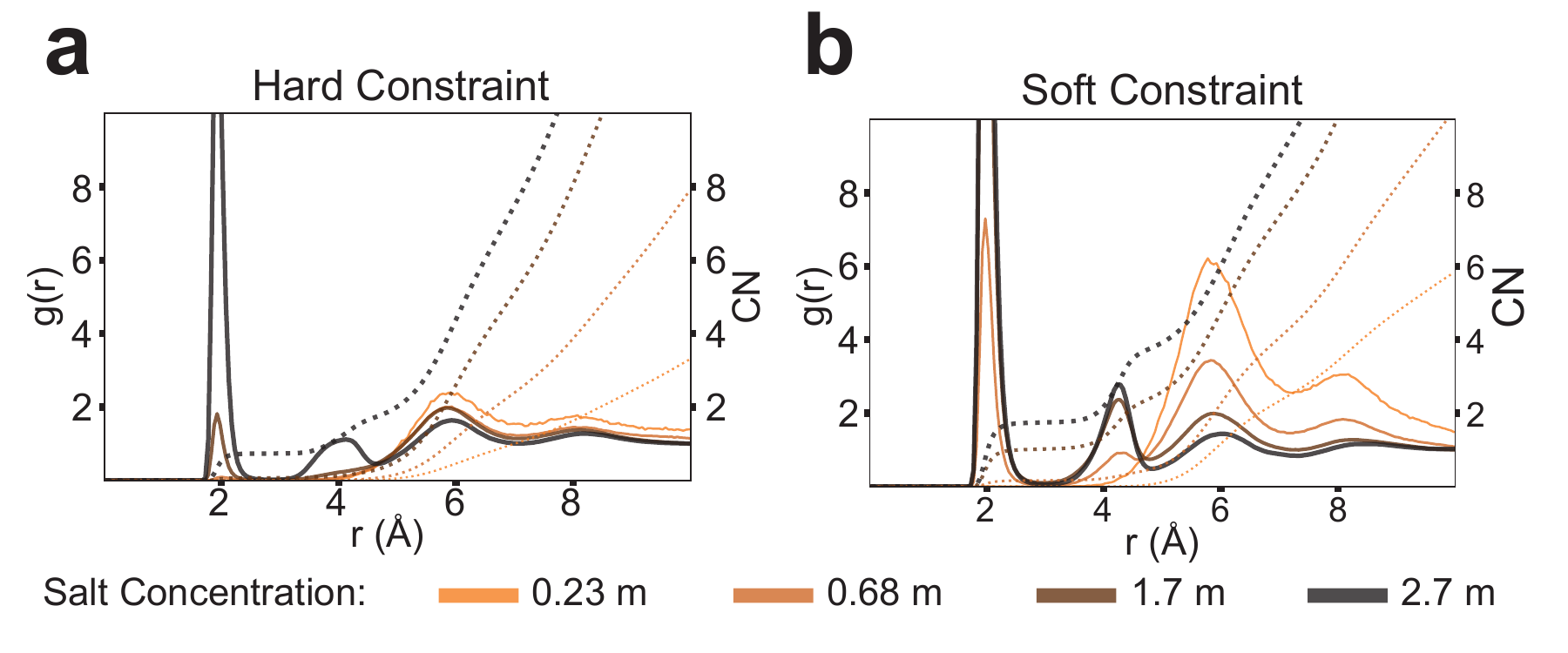}
  \caption{Li–O (TFSI) radial distribution functions (left axis, solid line) and corresponding coordination numbers (right axis, dashed line) at multiple molalities.
    a) Hard-constraint variant (b) Soft-constraint variant.}
  \label{fig:rdf-li-o}
\end{figure}

To further probe the local structure, we computed RDFs between Li-ions and TFSI oxygen atoms. As shown in Figure~\ref{fig:rdf-li-o} (b), the soft constraint algorithm RDF exhibits sharp peaks around 2~\AA\ and 4~\AA\ whose intensities increase with salt concentration and a broader peak at 6~\AA\ which diminishes at higher salt concentration. The overall RDF structure resembles that found in non-polarizable simulations discussed later in the text.

In contrast, the hard constraint algorithm (Fig.~\ref{fig:rdf-li-o}(a)) shows more moderate short-range structure at low concentrations, with 2~\AA\ and 4~\AA\ peaks becoming pronounced only at the highest salt concentrations. Both models exhibit qualitative agreement with experimental structural analysis.\cite{Mao2000} However, the hard-constraint algorithm shows closer consistency with reported coordination trends at low concentration due to the reduced intensity of the shortest-distance peak, suggesting a more moderate description of Li-TFSI contact pairing under dilute conditions.

\subsection{Kinetic Analysis}

\begin{figure}[H]

  \includegraphics[width=\linewidth]{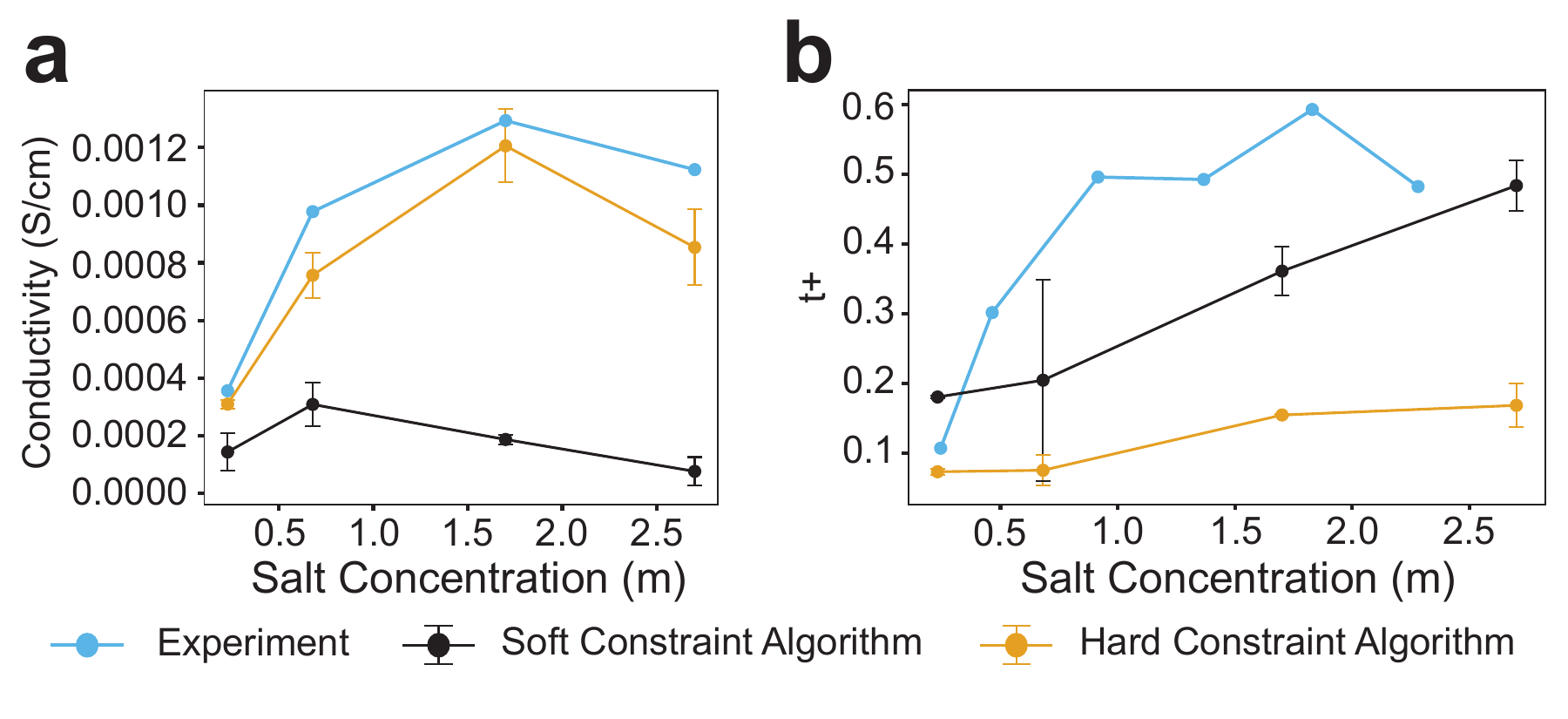}
  \caption{
    Comparison of (a) Cluster-Nernst-Einstein conductivities  (b) cation transference numbers as a function of molality at 353 K for both algorithms and experiment. Experimental conductivities from ref.~\citenum{Ferry1994} at 353~K, the same temperature as the simulations; experimental transference numbers from ref.~\citenum{Villaluenga2018} at 363~K, in the barycentric frame of ref.~\citenum{Shao2022}. Error bars are $\pm 2\sigma$ over two independent simulations.}
  \label{fig:cne-pol-cond}
\end{figure}

As shown in Fig.~\ref{fig:cne-pol-cond}(a), the hard constraint model exhibits a better agreement with experimental conductivity trends\cite{Ferry1994} with a mean absolute percentage error of 17\%. In contrast, the soft constraint model fails to reproduce the correct concentration dependence, predicting an incorrect location of the conductivity maximum and yielding a mean absolute percentage error of 77\%.

The differing trends are consistent with the structural analysis discussed above. The soft-constraint model predicts stronger Li-TFSI coordination at low and intermediate concentrations, which likely suppresses effective charge transport by stabilizing contact ion pairs and large clusters. The hard-constraint model, which exhibits more moderate short-range coordination, produces transport trends that better align with experiment.

Both polarizable models reproduce the increase of the cation transference number with salt concentration observed experimentally (Fig.~\ref{fig:cne-pol-cond}b). The experimental values are the transference numbers $t_+^0$ measured by Villaluenga \latin{et al.} at 363~K for 5~kg\,mol$^{-1}$ PEO,\cite{Villaluenga2018} which are defined relative to the solvent velocity and become negative near $r = 0.14$; we use their conversion to the barycentric reference frame by Shao \latin{et al.},\cite{Shao2022} the frame appropriate for comparison with MD, in which they remain positive. The hard-constraint model is closest to experiment in the dilute limit (0.07 versus 0.11 at 0.23~molal) but plateaus near 0.17 above 1.7~molal, where experiment reaches $\approx$0.5. The soft-constraint model overestimates the dilute-limit value (0.18) but approaches the experimental magnitude at high concentration (0.48 versus 0.45 at 2.7~molal). We attribute the high-concentration plateau of the hard-constraint model to the onset of strong Li--TFSI contact pairing at 2.7~molal (Fig.~\ref{fig:rdf-li-o}a), which immobilizes Li in neutral or anion-rich clusters.

\subsection{Analysis of Charge Scaling}

Charge-scaled fixed-charge models are the standard alternative to explicit polarization, so we use them as the control for our central claim. To this end, we fitted a non-polarizable force field within the same framework. For this model, the training loss was modified by removing constraints on equilibrium bond and angle parameters, consistent with previous work\cite{Leon2025} while the Lennard-Jones terms were kept fixed. The system sizes and simulation protocols were chosen to closely match those used for the polarizable version.

The resulting parity plots are shown in Supporting Information Figure~\ref{fig:class 1 partiy}. The non-polarizable model exhibits lower MAE values for energy and force compared to the polarizable models, and a dipole MAE between those of the soft- and hard-constraint variants. The predicted dipoles are qualitatively similar to those obtained from the soft-constraint Drude model. We attribute the higher parity errors of the polarizable models to the Drude degrees of freedom and dipole constraint acting as a regularizer, restricting how freely the bonded and charge parameters can fit the potential-energy surface.

The simulated densities for charge-scaling factors of 0.70, 0.75, and 0.80 are reported in Table \ref{tab:density_scaling_no_pol}. The densities are comparable in magnitude and concentration dependence to those obtained with the polarizable models, indicating that bulk packing is not strongly sensitive to the inclusion of explicit polarization.

\begin{figure}[H]
  \includegraphics[width=\linewidth]{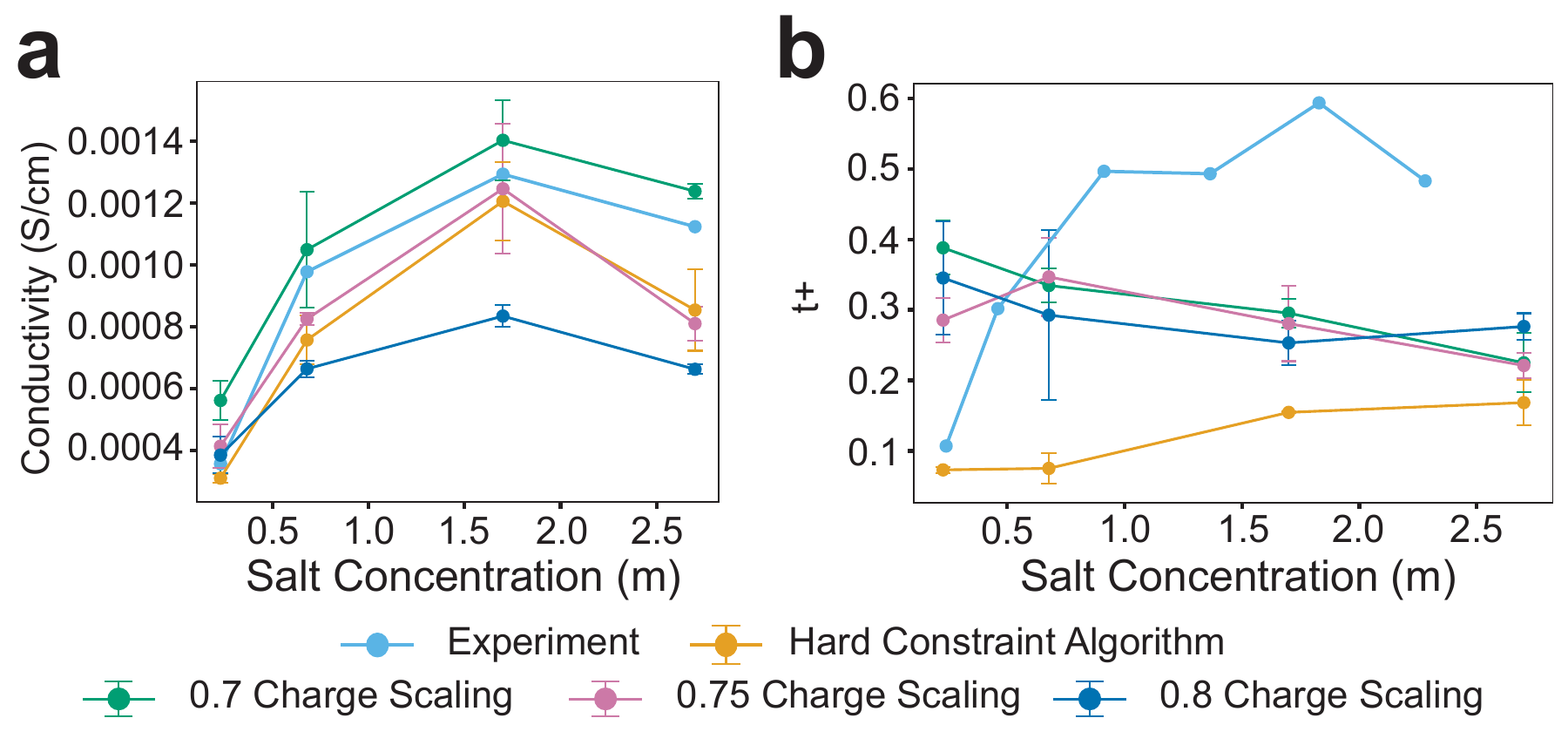}

  \caption{
    Comparison of a) cNE Conductivities  (b) cNE transference numbers as a function of molality at 353 K for different scalings, hard constraint algorithm, and experiment. Experimental data as in Fig.~\ref{fig:cne-pol-cond} (conductivities at 353~K; transference numbers at 363~K, barycentric frame). Error bars are $\pm 2\sigma$ over two independent simulations.}
  \label{fig:cne-non-pol-cond}
\end{figure}

We next computed cNE conductivities and transference numbers for salt charge scaling factors of 0.70, 0.75, and 0.80, consistent with common literature practice.\cite{Liang2024,Bedrov2019,Gudla2020} As shown in Figure \ref{fig:cne-non-pol-cond}(a), all scaling factors reproduce the qualitative concentration dependence of the experimental conductivity.\cite{Ferry1994} Table~\ref{tab:cond_errors} lists the mean absolute and mean absolute percentage errors of each model against the experimental conductivity. On absolute error the 0.70 scaling is closest, while on percentage error the hard-constraint polarizable model (17\%) is comparable to the best scaling factor (0.75, 16\%) and to 0.70 (21\%), whose dilute-limit conductivity is overestimated by 58\%. We stress that the scaling factor is selected against the experimental conductivity itself, whereas the polarizable model contains no empirically adjusted parameters. However, as shown in Fig.~\ref{fig:cne-non-pol-cond}b, the charge-scaled models overestimate the dilute-limit transference number by a factor of 2.7--3.6 and predict values that decrease (0.70 and 0.75) or stay flat (0.80) between 0.68 and 2.7 molal, opposite to experiment\cite{Villaluenga2018} and to both polarizable models. Charge scaling can therefore be tuned to reproduce the conductivity, but not the concentration dependence of cation transport.

\begin{table}[h!]
\centering
\caption{Simulated densities (g\,cm$^{-3}$) of PEO--LiTFSI non-polarizable systems at various molalities and charge--scaling factors at 353~K. Values are means of two independent simulations. Their standard deviation is below 0.01~g\,cm$^{-3}$ in all cases.}
\begin{tabular}{lcccc}
\hline
\textbf{Molality (mol\,kg$^{-1}$)} & 0.23 & 0.68 & 1.7 & 2.7 \\
\hline
\textbf{Scaling 0.70} & 1.05 & 1.11 & 1.23 & 1.32 \\[2pt]
\textbf{Scaling 0.75} & 1.05 & 1.13 & 1.24 & 1.34 \\[2pt]
\textbf{Scaling 0.80} & 1.06 & 1.13 & 1.25 & 1.34 \\
\hline
\end{tabular}
\label{tab:density_scaling_no_pol}
\end{table}

\begin{table}[h!]
\centering
\caption{Errors of the cNE conductivities with respect to experiment.\cite{Ferry1994}}
\begin{tabular}{lcc}
\hline
\textbf{Model} & \textbf{MAE ($10^{-4}$~S\,cm$^{-1}$)} & \textbf{MAPE (\%)} \\
\hline
Hard constraint & 1.56 & 17 \\
Soft constraint & 7.59 & 77 \\
Scaling 0.70    & 1.25 & 21 \\
Scaling 0.75    & 1.43 & 16 \\
Scaling 0.80    & 3.16 & 29 \\
\hline
\end{tabular}
\label{tab:cond_errors}
\end{table}

\begin{figure}
  \includegraphics[width=\linewidth]{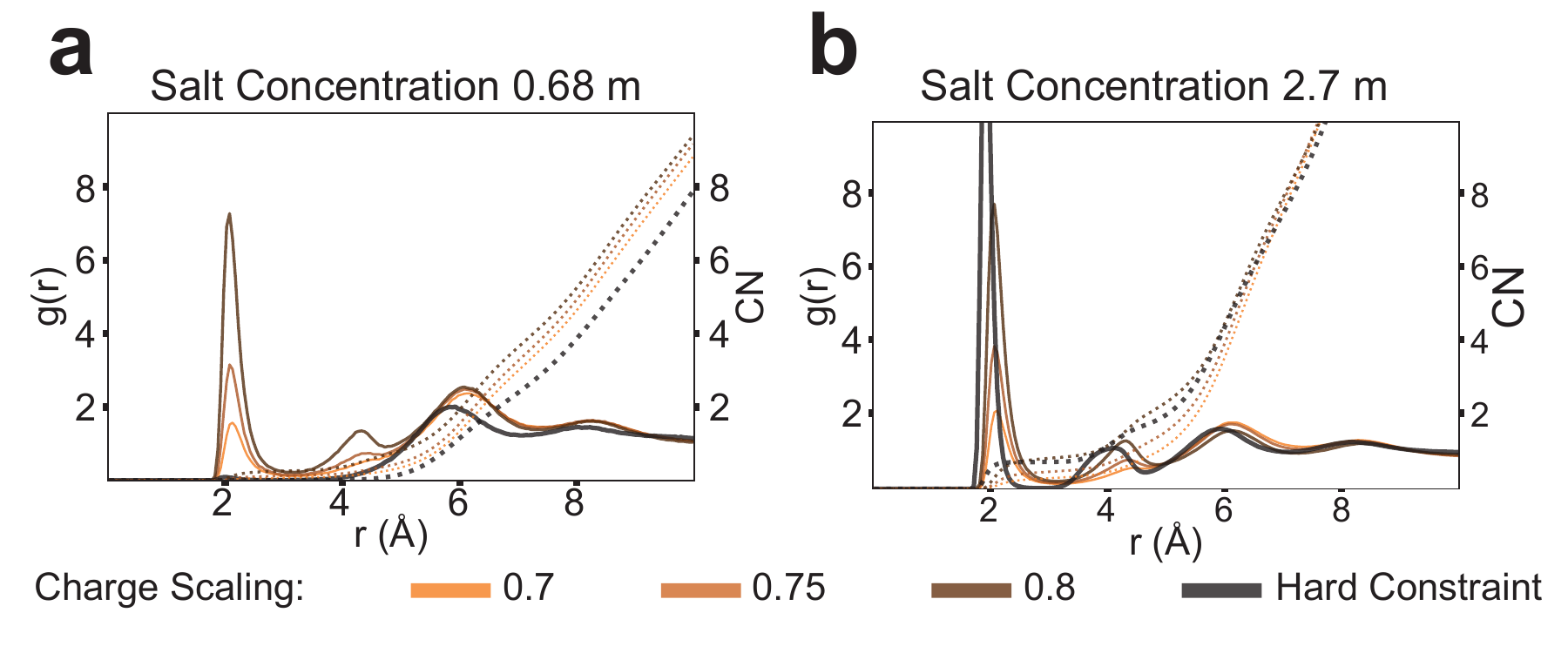}

  \caption{
    Li–O (TFSI) radial distribution functions (left axis, solid line) and coordination numbers (right axis, dashed line) at a) 0.68 m and b) 2.7 m}
  \label{fig:gr-non-pol}
\end{figure}

To elucidate the structural origin of these transport differences, we computed the Li–TFSI oxygen RDFs at 0.68 and 2.7 molal concentrations as shown in Figure \ref{fig:gr-non-pol}. At 0.68 molal, the non-polarizable models exhibit a pronounced peak at 2~\AA, which is absent in the polarizable case. We attribute this to overbinding of Li to TFSI in the absence of polarization, consistent with previous reports \cite{Liang2024}. At 2.7 molal, all models exhibit pronounced short-range structure. However, systematic differences remain. Imposing stronger charge scaling from 0.80 to 0.70 progressively shifts intensity from the 2~\AA\ peak toward longer distances at 4~\AA, indicating reduced contact ion pairing. We find that the polarizable model lies between 0.70 and 0.80 scaling cases, suggesting a more balanced description of short- and intermediate-range ion correlations.

We conclude that the addition of explicit polarization in this system serves to reduce overbinding at low salt concentrations and alters the balance between contact ion pairs and solvent-separated configurations. When considered alongside the transference number results, we find that while charge scaling can reproduce experimental conductivity trends, it does so without consistently reproducing the underlying structural motifs and correlated transport behavior reflected in the transference number.

\subsection{Computational Performance}

All molecular dynamics simulations were performed using OpenMM on a single NVIDIA Volta V100 GPU. For the 1.7~molal PEO–LiTFSI system (10,516 atoms; 12,372 particles including Drude particles), the classical non-polarizable force field achieved 165 ns/day, while the Drude polarizable force field achieved 138 ns/day, corresponding to an overhead of 1.20$\times$ at equal 1~fs timesteps ($\approx$2.4$\times$ relative to a classical run at the customary 2~fs). This represents a significant improvement over previous reports of 2–4$\times$ overhead for Drude simulations on GPU hardware.\cite{Huang2021} Modern GPU-accelerated OpenMM implementations have substantially reduced the computational penalty of explicit polarization.\cite{Eastman2024} The modest overhead demonstrates that polarizable force fields are now practical for long-timescale simulations of polymer electrolytes and may be feasible for large-scale screening studies.

\section{Conclusion}

We have demonstrated that AutoDrude enables automated, differentiable fitting of polarizable force fields capable of reproducing both structural and dynamic properties of PEO–LiTFSI electrolytes with agreement to experimental conductivity comparable to that of optimally charge-scaled non-polarizable models, obtained without empirical tuning, and with a concentration dependence of the cation transference number that fixed-charge models do not reproduce. By simultaneously enforcing consistency of energy, force, and dipole with DFT reference data, AutoDrude learns atom-specific polarizabilities, Thole damping factors, and bonded parameters that have traditionally required years of manual tuning. The resulting Drude-based potentials capture salt concentration dependent trends in density, coordination, and conductivity that remain challenging for fixed-charge models without charge scaling, consistent with an important role for explicit polarization in describing coupled ion–polymer dynamics. Of the two polarization strategies examined, we select the hard-constraint formulation because it is the only model, polarizable or not, that reproduces the conductivity (17\% mean absolute percentage error, comparable to the best charge-scaling factor), the dilute-limit transference number (0.07 versus 0.11 experimentally) and the absence of Li--TFSI contact pairs at low concentration together. The soft-constraint variant is closer to experiment in the transference number at high concentration (0.48 versus 0.45 at 2.7~molal) and in the ordering of Li--TFSI cluster energetics (Fig.~\ref{fig:tfsi-coord-dih}a--b), but underestimates the conductivity by up to an order of magnitude.

Our results further demonstrate that accurate reproduction of quantum mechanical energies and forces alone does not guarantee correct emergent transport behavior, as previously reported for fixed-charge potentials.\cite{Leon2025} Charge-scaled non-polarizable force fields may reproduce certain experimental observables such as bulk conductivity trends, yet fail to consistently reproduce the underlying coordination environments and correlated ion motion reflected in cation transference numbers. These findings point to a role for explicit polarization in ion-polymer coupling that empirical charge scaling does not reproduce.

Although AutoDrude represents a significant advancement toward data-driven polarizable force-field development, several challenges remain. The present implementation parametrizes a specific system rather than transferable families of electrolytes (e.g., alternative anions), and does not yet provide an automated strategy for identifying polarizable sites or leveraging existing polarizable parameter libraries. Incorporating lone-pairs and anisotropic polarization terms common in advanced CHARMM parametrizations could further enhance its expressive power. Finally, larger systems and longer trajectories would allow Onsager transport coefficients to be converged directly and the system-size dependence of the cNE estimates to be assessed. Overall, AutoDrude establishes a scalable and differentiable framework for systematic polarizable force-field development and provides a pathway toward physically consistent, computationally tractable modeling of complex electrochemical materials.

\begin{acknowledgement}

This work was supported by the Toyota Research Institute and by the Energy Storage Research Alliance ”ESRA” (DE-AC02-06CH11357), an Energy Innovation Hub funded by the U.S. Department of Energy, Office of Science, Basic Energy Sciences. The authors acknowledge the MIT SuperCloud and Lincoln Laboratory Supercomputing Center for HPC resources as well as computational resources of the National Energy Research Scientific Computing Center (NERSC). We also acknowledge helpful discussions with Pablo Leon and Jurgis Ruza. AL acknowledges Giselle Chiprez-Ramirez for help with figure design. 
\\
\\
The code will be made available upon journal publication.

\end{acknowledgement}

%%%%%%%%%%%%%%%%%%%%%%%%%%%%%%%%%%%%%%%%%%%%%%%%%%%%%%%%%%%%%%%%%%%%%
%% The same is true for Supporting Information, which should use the
%% suppinfo environment.
%%%%%%%%%%%%%%%%%%%%%%%%%%%%%%%%%%%%%%%%%%%%%%%%%%%%%%%%%%%%%%%%%%%%%
\begin{suppinfo}

\section{Supporting Information}

\subsection{S1. Dataset and Parameter Initialization}
The dataset used is taken from a previous study.\cite{Leon2025} The generation method of the geometries is summarized in Figure~\ref{fig:Dataset}. Full protocols are given in the Supporting Information of ref.~\citenum{Leon2025}. Briefly, bond, angle, and dihedral scans and normal-mode (Hessian) displacements of triglyme were generated with GFN1-xTB; classical MD frames were sampled in the gas phase with fitted class II (PCFF-form; triglyme and Li--TFSI, NVT) and class I (OPLS-form; tetraglyme, NVE) force fields; MLIP MD frames were sampled with a neural-network potential trained on BP86-D3 data; and Li--TFSI geometries were obtained from DFT optimizations and displacements around those minima, seeded from GFN2-xTB structures, pose guesses, and MLIP MD frames. All energies, forces, CHELPG charges and dipoles were computed with ORCA at the $\omega$B97X-D3/def2-TZVP level. Geometries were centred at their centroid before dipoles were evaluated.

\begin{figure}[H]
  \centering
  \includegraphics[width=0.75\linewidth]{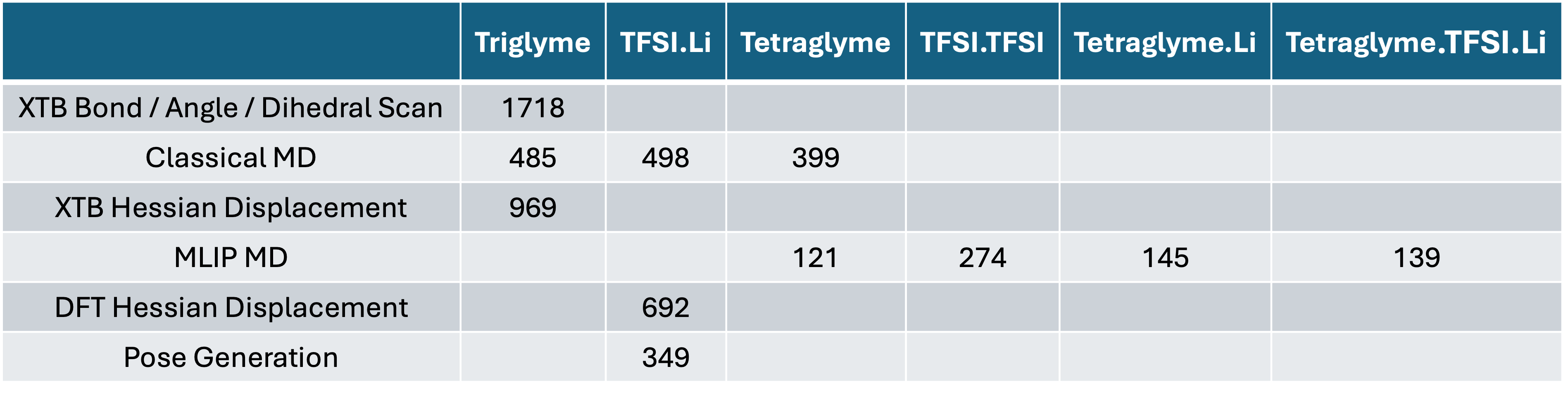}
  \caption{Dataset used for training.}
  \label{fig:Dataset}
\end{figure}

Bond and angle equilibrium parameters are set to the average over the DFT frames. As the force constants are not readily available we set them to a sensible order of magnitude guess. Dihedral force constants and equilibrium values are set to 0.01 kcal/mol and 0 radians respectively. Charges are assigned as an average for the given atom over the DFT frames. We fix Lennard Jones parameters from CHARMM polarizable force field files as the DFT level of theory employed here is not sufficient to fit them.\cite{Bukowski2013,Baker2010b,MacKerell2016} Lorentz–Berthelot mixing and standard CHARMM cutoffs/switching were used. 

Drude parameters are set such that their bond constants are 1000 kcal/mol/$\text{\r{A}}^2$ while fixing the polarizability to the closest matching values from the parameter files of the CHARMM polarizable force field. Thole screening is set to 1.3. 

Two independent sets of polarization and Thole parameters are used to parameterize the intramolecular and intermolecular interactions involving drudes. The intramolecular parameters handle the bonded dipole screening and the intermolecular parameters serve as the NBTHOLE screening in the polarizable CHARMM force field. The atom typing is in accordance with explicit screening of 1-2 and 1-3 neighbors and 1-4 neighbors remain unscreened. The intermolecular parameters are attached to a given atom type and these parameters are combined to produce the non-bonded screening. 

\subsection{S2. Minimization convergence}

In order to validate our constrained optimization algorithm we show the results of our drude position optimization algorithm for the trained force field discussed in the main paper. We report RMS drude force, RMS dipole residual with respect to the DFT dipole, and the magnitude of the drude position update. The results are shown for all the frames in our dataset for tetraglyme, a cluster of two TFSI, and a tetraglyme-TFSI-lithium complex. We start off with three warm up gradient descent steps (shaded region) and show results for three and five constrained update steps. As can be seen in the panels the constrained algorithm increases the end force on the drudes due to the alignment with the molecular dipole. The force plateaus after the constrained optimization starts, shifting the position of the drudes and minimizing the error to the DFT dipole. None of the metrics presented change appreciably between 3 and 5 constrained optimization steps. 

\begin{figure}[H]
  \centering
  \setlength{\tabcolsep}{2pt} % horizontal gap between cells
  \begin{tabular}{@{}ccc@{}}
    \includegraphics[width=0.32\linewidth]{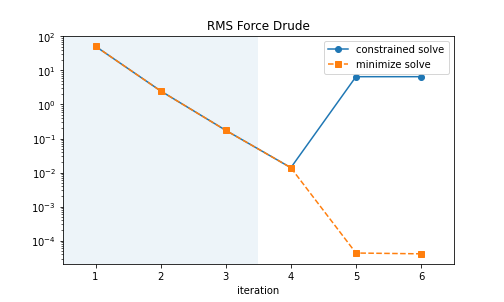} &
    \includegraphics[width=0.32\linewidth]{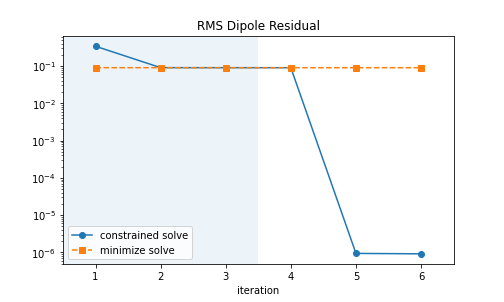} &
    \includegraphics[width=0.32\linewidth]{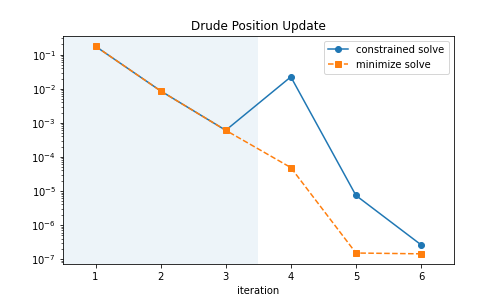} \\
    {\footnotesize (a) RMS Force on Drude} &
    {\footnotesize (b) RMS Dipole Residual} &
    {\footnotesize (c) Drude Position Update}
  \end{tabular}
  \caption{Convergence metrics for tetraglyme with 3 constrained steps}
  \label{fig:Peo 3-3}
\end{figure}

\begin{figure}[H]
  \centering
  \setlength{\tabcolsep}{2pt} % horizontal gap between cells
  \begin{tabular}{@{}ccc@{}}
    \includegraphics[width=0.32\linewidth]{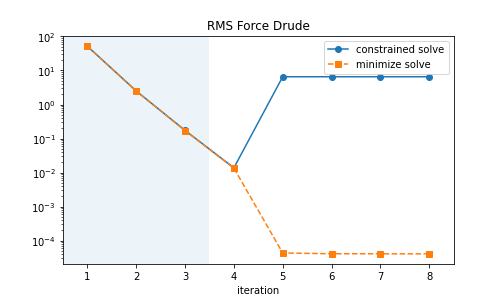} &
    \includegraphics[width=0.32\linewidth]{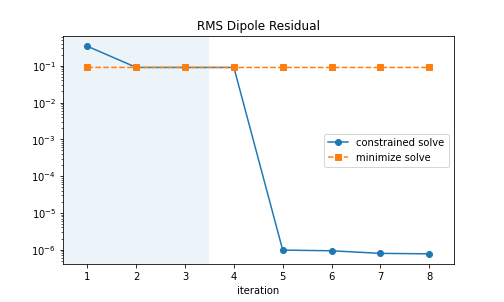} &
    \includegraphics[width=0.32\linewidth]{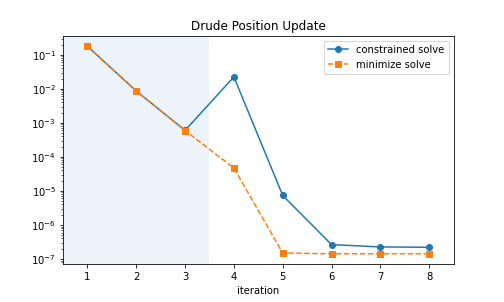} \\
    {\footnotesize (a) RMS Force on Drude} &
    {\footnotesize (b) RMS Dipole Residual} &
    {\footnotesize (c) Drude Position Update}
  \end{tabular}
  \caption{Convergence metrics for tetraglyme with 5 constrained steps}
  \label{fig:Peo 3-5}
\end{figure}

\begin{figure}[H]
  \centering
  \setlength{\tabcolsep}{2pt} % horizontal gap between cells
  \begin{tabular}{@{}ccc@{}}
    \includegraphics[width=0.32\linewidth]{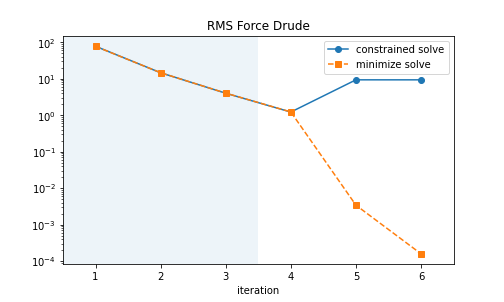} &
    \includegraphics[width=0.32\linewidth]{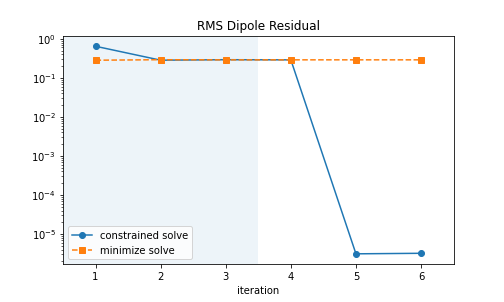} &
    \includegraphics[width=0.32\linewidth]{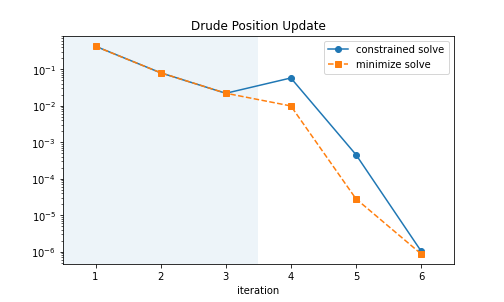} \\
    {\footnotesize (a) RMS Force on Drude} &
    {\footnotesize (b) RMS Dipole Residual} &
    {\footnotesize (c) Drude Position Update}
  \end{tabular}
  \caption{Convergence metrics for TFSI cluster with 3 constrained steps}
  \label{fig:TFSI 3-3}
\end{figure}

\begin{figure}[H]
  \centering
  \setlength{\tabcolsep}{2pt} % horizontal gap between cells
  \begin{tabular}{@{}ccc@{}}
    \includegraphics[width=0.32\linewidth]{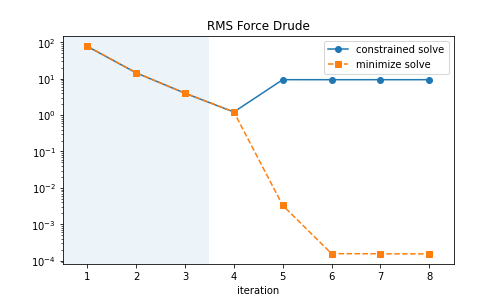} &
    \includegraphics[width=0.32\linewidth]{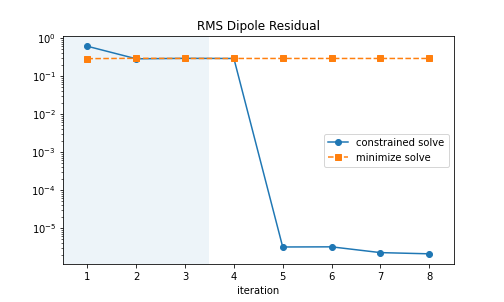} &
    \includegraphics[width=0.32\linewidth]{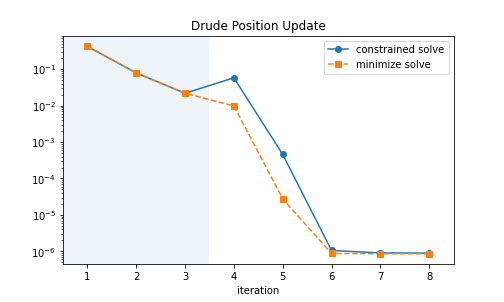} \\
    {\footnotesize (a) RMS Force on Drude} &
    {\footnotesize (b) RMS Dipole Residual} &
    {\footnotesize (c) Drude Position Update}
  \end{tabular}
  \caption{Convergence metrics for TFSI cluster with 5 constrained steps}
  \label{fig:TFSI 3-5}
\end{figure}

\begin{figure}[H]
  \centering
  \setlength{\tabcolsep}{2pt} % horizontal gap between cells
  \begin{tabular}{@{}ccc@{}}
    \includegraphics[width=0.32\linewidth]{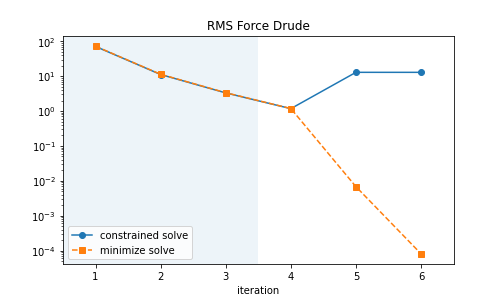} &
    \includegraphics[width=0.32\linewidth]{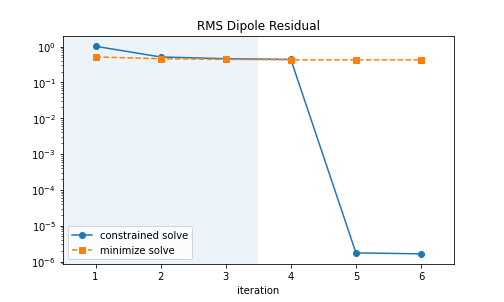} &
    \includegraphics[width=0.32\linewidth]{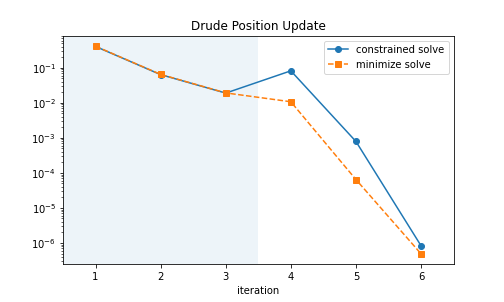} \\
    {\footnotesize (a) RMS Force on Drude} &
    {\footnotesize (b) RMS Dipole Residual} &
    {\footnotesize (c) Drude Position Update}
  \end{tabular}
  \caption{Convergence metrics for tetraglyme-TFSI-lithium cluster with 3 constrained steps}
  \label{fig:PEO TFSI Li 3-3}
\end{figure}

\begin{figure}[H]
  \centering
  \setlength{\tabcolsep}{2pt} % horizontal gap between cells
  \begin{tabular}{@{}ccc@{}}
    \includegraphics[width=0.32\linewidth]{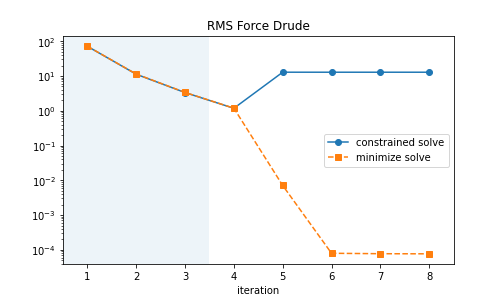} &
    \includegraphics[width=0.32\linewidth]{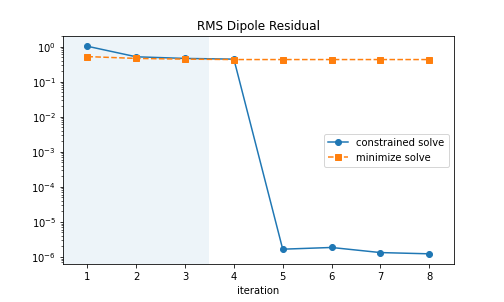} &
    \includegraphics[width=0.32\linewidth]{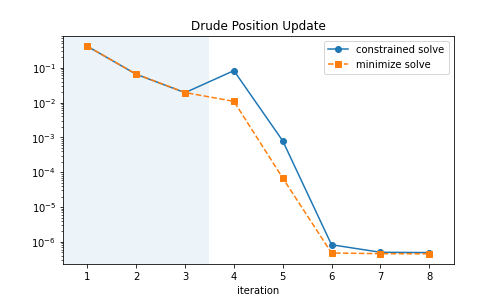} \\
    {\footnotesize (a) RMS Force on Drude} &
    {\footnotesize (b) RMS Dipole Residual} &
    {\footnotesize (c) Drude Position Update}
  \end{tabular}
  \caption{Convergence metrics for tetraglyme-TFSI-lithium cluster with 5 constrained steps}
  \label{fig:PEO TFSI Li 3-5}
\end{figure}

\subsection{S3. Training}

Following prior literature reports that bond and angle coefficients were difficult optimization targets we perform parameter optimization on the square root of the bond and angle constants. An Adam optimizer was used with an initial learning rate of 0.01 and ReduceLROnPlateau was used with a final learning rate of $10^{-4}$. All training is performed for a total of 2000 epochs.

\subsubsection{Polarization removed on H}
\begin{figure}[H]
  \centering
  % left panel
  \begin{minipage}[t]{0.48\linewidth}
    \centering
    \includegraphics[width=\linewidth]{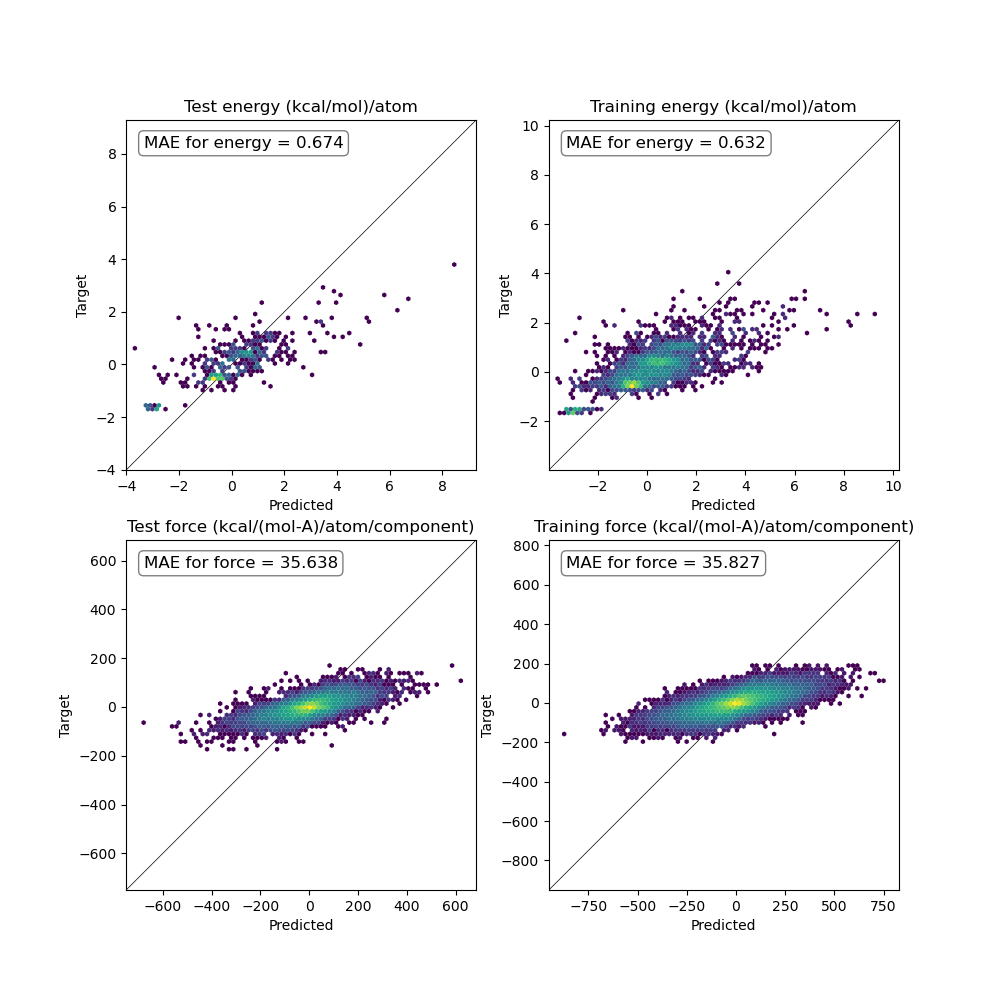}
    {\footnotesize (a) Hard constraint}
  \end{minipage}\hfill
  % right panel
  \begin{minipage}[t]{0.48\linewidth}
    \centering
    \includegraphics[width=\linewidth]{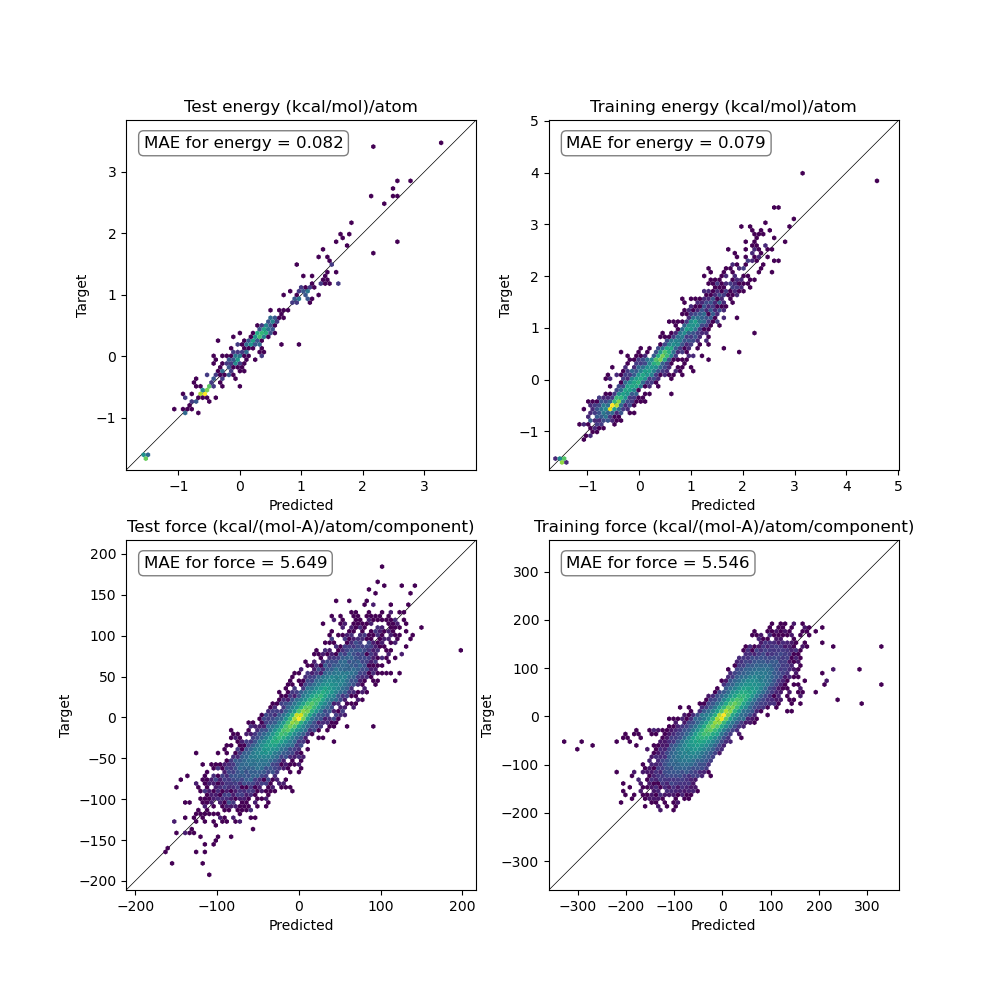}
    {\footnotesize (b) Soft constraint}
  \end{minipage}

  \caption{Parity plots for the train and test sets for each algorithm version.
    a) Hard-constraint variant (b) Soft-constraint variant.}
  \label{fig:H parity}
\end{figure}

For this polarization case it can be seen that the hard constraint algorithm produces large errors in energies and forces. The soft constraint algorithm produces satisfactory training plots but fits negative Thole values on TFSI sulfurs and some of PEO carbons.

\subsubsection{Polarization removed on H and C}

\begin{figure}[H]
  \centering
  % left panel
  \begin{minipage}[t]{0.48\linewidth}
    \centering
    \includegraphics[width=\linewidth]{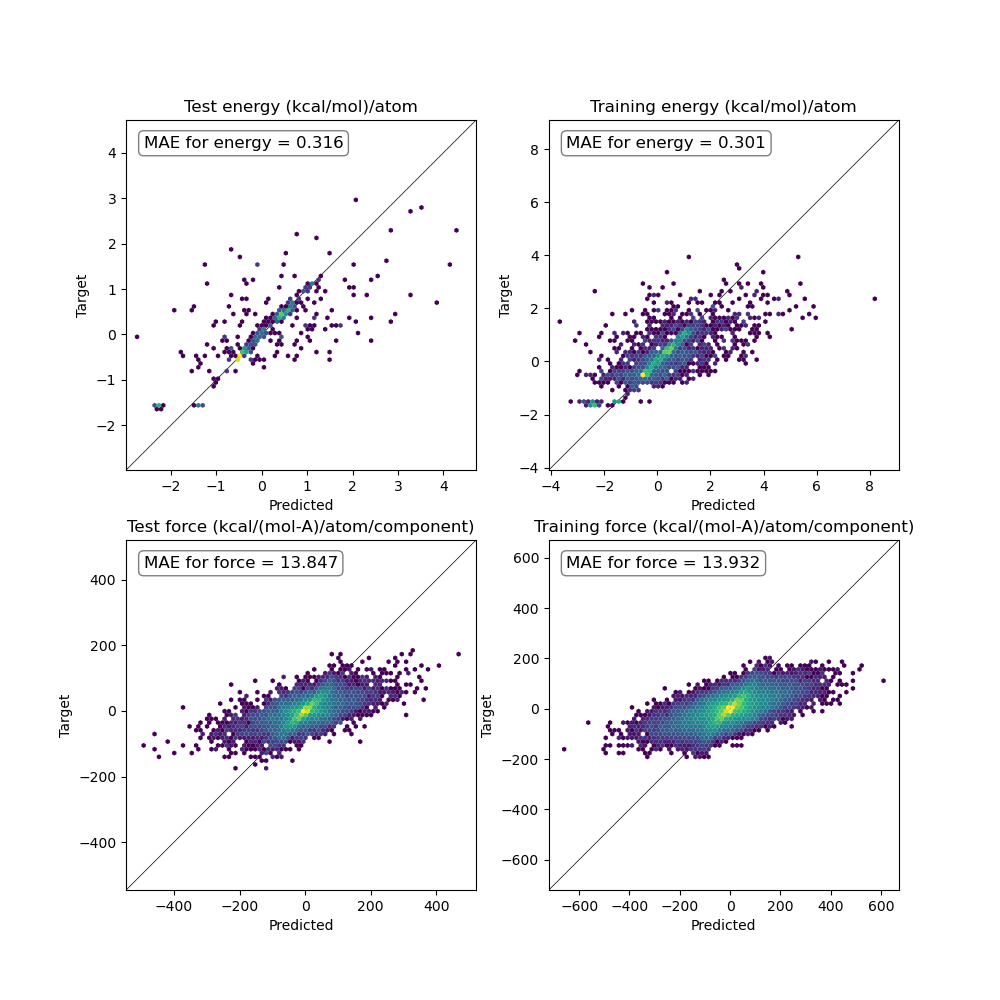}
    {\footnotesize (a) Hard constraint}
  \end{minipage}\hfill
  % right panel
  \begin{minipage}[t]{0.48\linewidth}
    \centering
    \includegraphics[width=\linewidth]{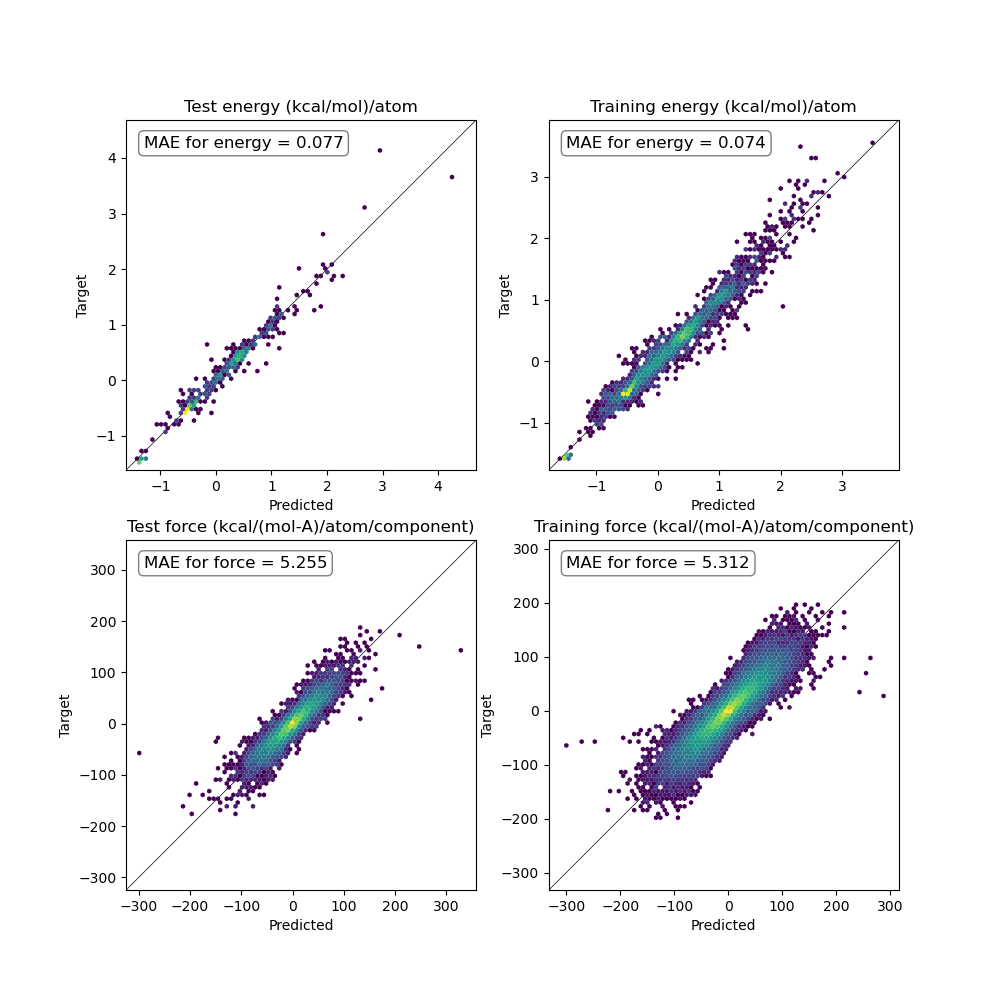}
    {\footnotesize (b) Soft constraint}
  \end{minipage}

  \caption{Parity plots for the train and test sets for each algorithm version.
    a) Hard-constraint variant (b) Soft-constraint variant.}
  \label{fig:H C parity}
\end{figure}

The hard constraint algorithm presents large errors in energies and forces. Soft constraint algorithm has negative Thole values on TFSI sulfur and oxygen.

\subsubsection{Polarization removed on H, C, and S}

\begin{figure}[H]
  \centering
  % left panel
  \begin{minipage}[t]{0.48\linewidth}
    \centering
    \includegraphics[width=\linewidth]{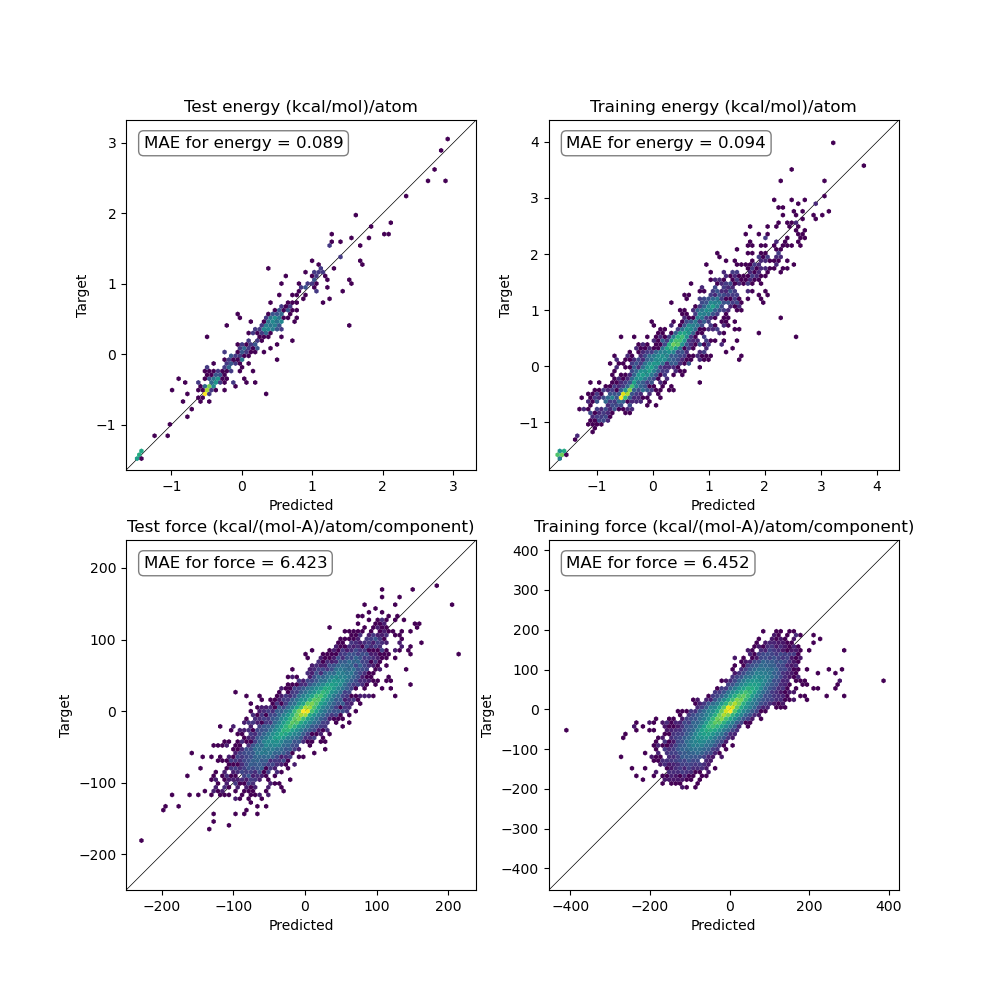}
    {\footnotesize (a) Hard constraint}
  \end{minipage}\hfill
  % right panel
  \begin{minipage}[t]{0.48\linewidth}
    \centering
    \includegraphics[width=\linewidth]{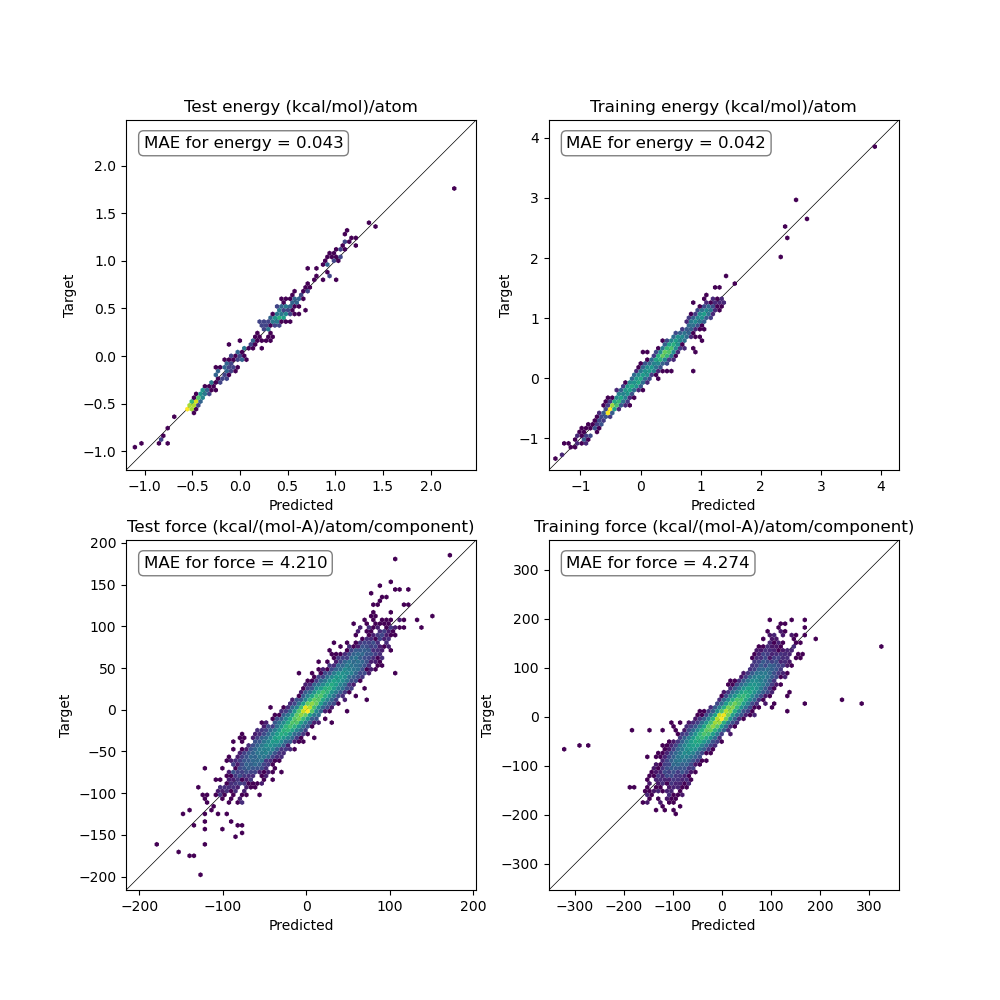}
    {\footnotesize (b) Soft constraint}
  \end{minipage}

  \caption{Parity plots for the train and test sets for each algorithm version.
    a) Hard-constraint variant (b) Soft-constraint variant.}
  \label{fig:H C S parity}
\end{figure}

\begin{figure}[H]
  \centering
  % left panel
  \begin{minipage}[t]{0.48\linewidth}
    \centering
    \includegraphics[width=\linewidth]{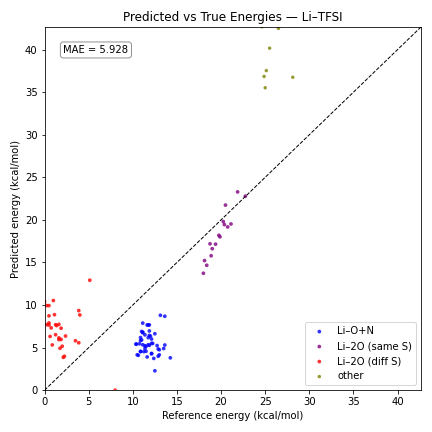}
    {\footnotesize (a) Hard constraint}
  \end{minipage}\hfill
  % right panel
  \begin{minipage}[t]{0.48\linewidth}
    \centering
    \includegraphics[width=\linewidth]{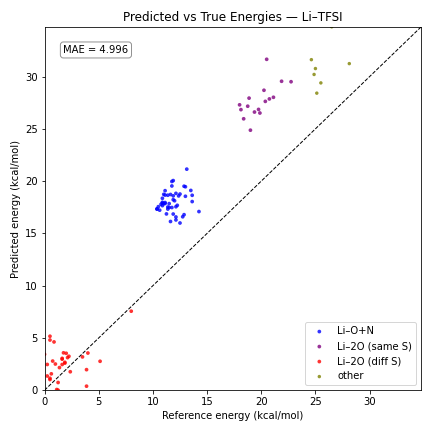}
    {\footnotesize (b) Soft constraint}
  \end{minipage}

  \caption{TFSI-Li parity plots based on Li coordination environment.
    (a) and (b) predicted versus DFT relative energies, $E - E_\text{min}$, of 103 Li--TFSI configurations not included in training. Coordination environments are based on the two nearest neighbors of the lithium-ion. Oxygen atoms bound to different sulfur centers within TFSI are distinguished from those bound
to the same sulfur.}
  \label{fig:H C S coord}
\end{figure}

Both algorithms produce satisfactory parity plots as well as reasonable parities between coordination environments. The trends can be seen to be opposite between the two algorithms. However, the intramolecular Thole factor of F collapses to $\leq 10^{-3}$ in both variants, and both force fields lead to salt segregation in MD tests.

\subsection{S4. Dihedrals}

\begin{figure}[H]
  \centering
  \includegraphics[width=0.75\linewidth]{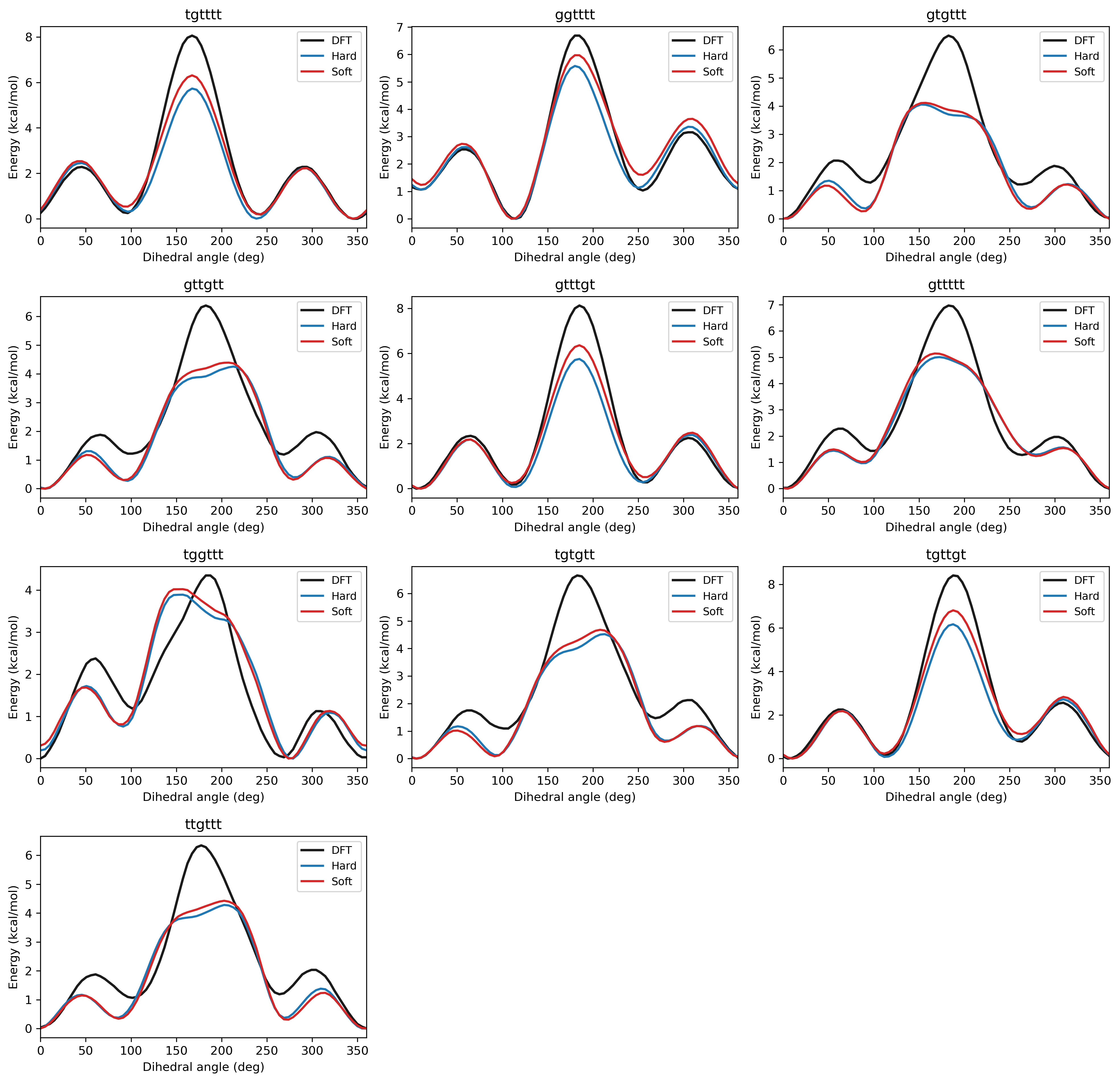}
  \caption{Dihedral scans of diglyme comparing DFT and fitted force-field
energy profiles. The characters “t” and “g” denote trans or gauche conformations, with
the right-most character indicating the scanned dihedral.}
  \label{fig:alldih}
\end{figure}

\subsection{S5. Non-polarizable force field}
\begin{figure}[H]
  \includegraphics[width=\linewidth]{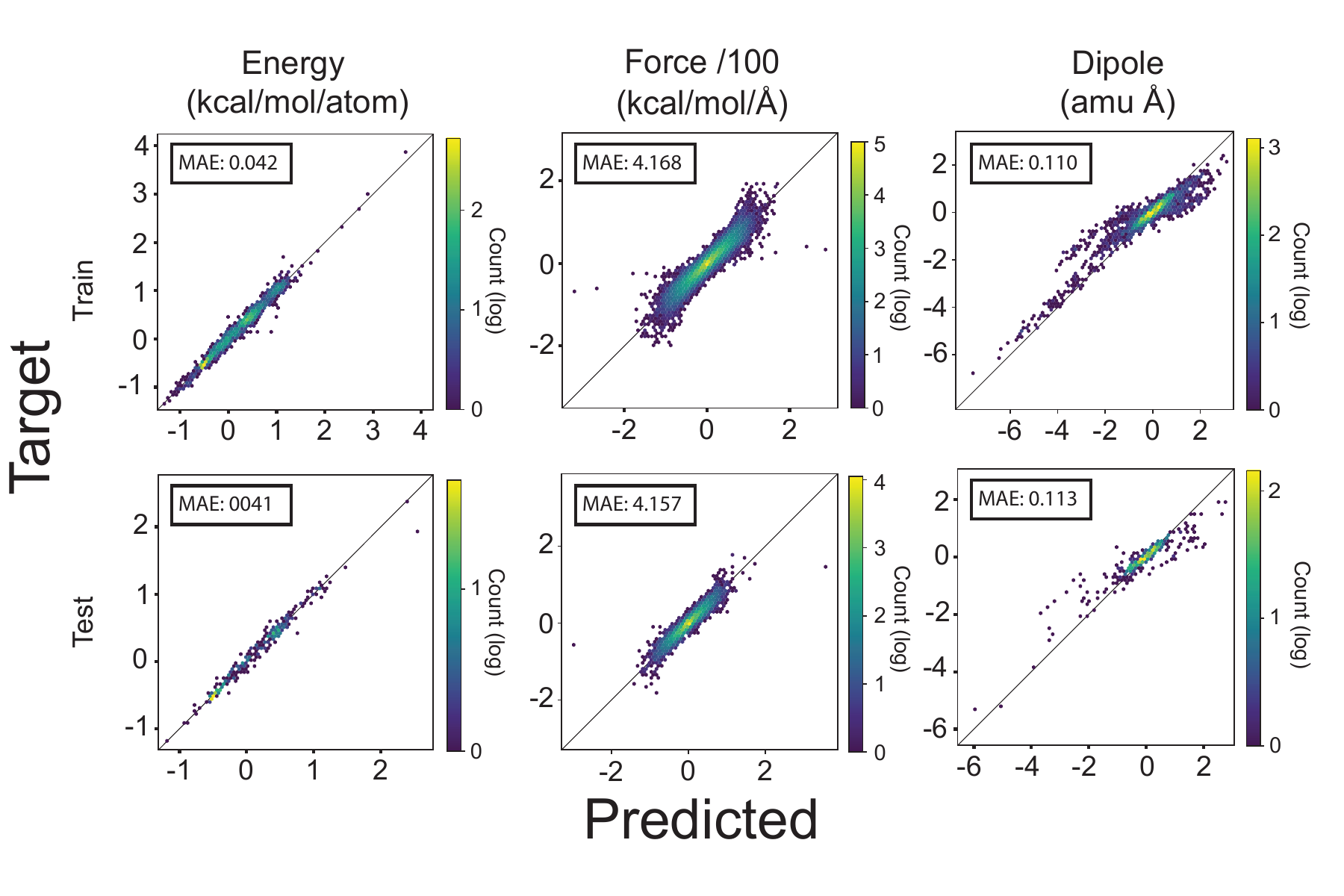}
  \caption{Parity plots for training and test sets of non-polarizable (classical) force field. Predicted values are shown on the x-axis and DFT reference values on the y-axis.}
  \label{fig:class 1 partiy}
\end{figure}

\begin{figure}[H]
  \centering
  \includegraphics[width=\linewidth]{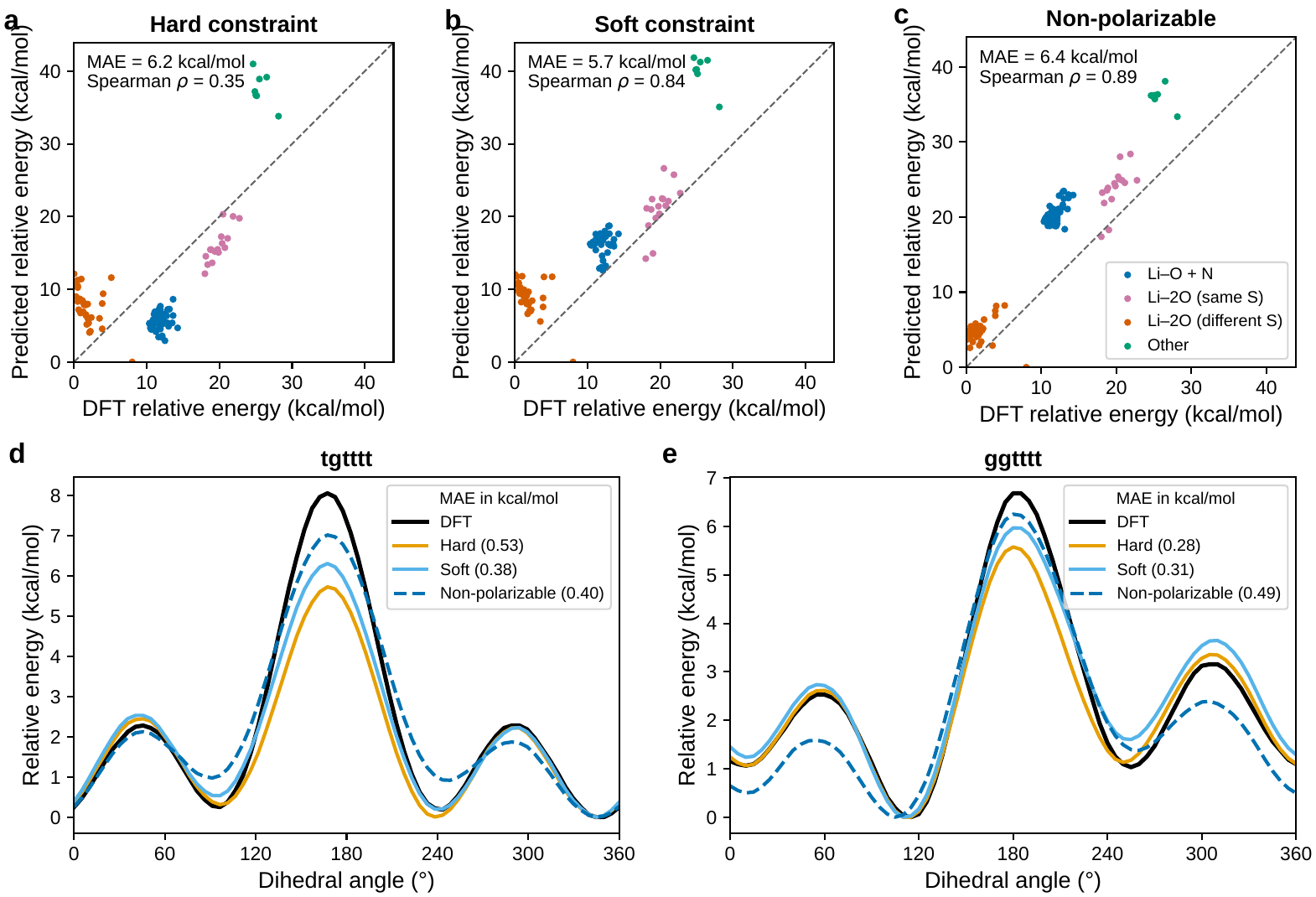}
  \caption{The comparison of Fig.~\ref{fig:tfsi-coord-dih} including the non-polarizable model. (a--c) Predicted versus DFT relative energies, $E - E_\text{min}$, of the 103 Li--TFSI configurations for the (a) hard-constraint, (b) soft-constraint, and (c) non-polarizable models; mean absolute errors and Spearman rank correlations are given in each panel. (d--e) Dihedral scans of diglyme for tgtttt and ggtttt.}
  \label{fig:classical-poses-scans}
\end{figure}

\subsection{S6. Experimental transference numbers}
Villaluenga \latin{et al.} report $t_+^0$ for 5~kg\,mol$^{-1}$ PEO/LiTFSI at 363~K, obtained from concentrated-solution theory in the solvent-velocity reference frame.\cite{Villaluenga2018} These values are converted according to Shao \latin{et al.} to the barycentric frame with the Woolf--Harris relation\cite{Shao2022}
\begin{equation}
\omega_0\, t_+^0 = t_+^{M} - \omega_-,
\end{equation}
where $\omega_0$ and $\omega_-$ are the mass fractions of polymer and anion. Figures~\ref{fig:cne-pol-cond}b and \ref{fig:cne-non-pol-cond}b show the barycentric values $t_+^{M}$.

\end{suppinfo}

%%%%%%%%%%%%%%%%%%%%%%%%%%%%%%%%%%%%%%%%%%%%%%%%%%%%%%%%%%%%%%%%%%%%%
%% The appropriate \bibliography command should be placed here.
%% Notice that the class file automatically sets \bibliographystyle
%% and also names the section correctly.
%%%%%%%%%%%%%%%%%%%%%%%%%%%%%%%%%%%%%%%%%%%%%%%%%%%%%%%%%%%%%%%%%%%%%
\bibliography{paper_main}

\end{document}